\documentclass[numberedappendix,twocolumn,twocolappendix]{openjournal}

\usepackage{latexsym}
\usepackage{graphicx}
\usepackage{amssymb}
\usepackage{longtable}
\usepackage{epsf}
\usepackage{amsmath}
\usepackage{graphicx}
\usepackage{gensymb}

\usepackage{hyperref}
\hypersetup{colorlinks=true,linkcolor=blue,citecolor=blue,filecolor=blue,urlcolor=blue}
\usepackage{cleveref}
\usepackage{bm}

\usepackage{natbib}

\newcommand{\Mstar}{M_{\star}}
\newcommand{\Mstari}{M_{\star,i}}
\newcommand{\msun}{M_{\odot}}
\newcommand{\mhalo}{M_{\rm halo}}
\newcommand{\mpeak}{M_{\rm peak}}
\newcommand{\mh}{m_{\rm h}}
\newcommand{\ms}{m_{\star}}
\newcommand{\mcrit}{m_{\rm c}}

\newcommand{\Mstarlo}{M_{\star}^{\rm lo}}
\newcommand{\Mstarhi}{M_{\star}^{\rm hi}}

\newcommand{\beq}{\begin{eqnarray}}
\newcommand{\eeq}{\end{eqnarray}}
\newcommand{\ben}{\begin{enumerate}}
\newcommand{\een}{\end{enumerate}}

\usepackage{xspace}
\newcommand{\shamnet}{{SHAMNet}\xspace}

\newcommand{\smhm}{\langle\Mstar\vert\mpeak\rangle_{\rm med}}
\newcommand{\smhmavg}{\langle\Mstar\vert\mpeak\rangle}

\newcommand{\smhmavgprime}{\langle\Mstar'\vert\mpeak\rangle}

\newcommand{\smhmtheta}{\langle\Mstar\vert\mpeak;\theta\rangle_{\rm med}}
\newcommand{\fsham}{\mathcal{F}_{\rm SHAM}}
\newcommand{\psisham}{\psi_{\rm SHAM}}

\newcommand{\dd}{\rm d}
\newcommand{\xcrit}{x_{\rm c}}
\newcommand{\ds}{\Delta\Sigma}
\newcommand{\dsrp}{\Delta\Sigma(R)}
\newcommand{\rproj}{{\rm r_p}}
\newcommand{\mpp}{{\rm m_p}}
\newcommand{\phigal}{\phi_{\rm g}}
\newcommand{\phih}{\phi_{\rm h}}

\newcommand{\activation}{\Upsilon_{\rm act}}
\newcommand{\wproj}{w_{\rm p}}
\newcommand{\wprp}{\wproj(\rproj)}

\usepackage[table]{xcolor}

\definecolor{hpurple}{HTML}{7E16DF}
\definecolor{hgreen}{HTML}{008F0F}
\definecolor{horange}{HTML}{FFA301}

\begin{document}

\title{Differentiable Predictions for Large Scale Structure with \shamnet}
\shorttitle{\shamnet}

\author{Andrew P. Hearin$^{1,\star}$}
\author{Nesar Ramachandra$^{2,1}$}
\author{Matthew R. Becker$^1$}
\author{Joseph DeRose$^{3}$}

\affiliation{$^1$High Energy Physics Division, Argonne National Laboratory, 9700 South Cass Avenue, Lemont, IL 60439, USA}
\affiliation{$^2$Computational Science Division, Argonne National Laboratory, 9700 South Cass Avenue, Lemont, IL 60439, USA}
\affiliation{$^{3}$Lawrence Berkeley National Laboratory, 1 Cyclotron Road, Berkeley, CA, 94720 USA}
\thanks{$^{\star}$E-mail:ahearin@anl.gov}

\shortauthors{Hearin et al.}


\begin{abstract}
In simulation-based models of the galaxy--halo connection, theoretical predictions for galaxy clustering and lensing are typically made based on Monte Carlo realizations of a mock universe. In this paper, we use Subhalo Abundance Matching (SHAM) as a toy model to introduce an alternative to stochastic predictions based on mock population, demonstrating how to make simulation-based predictions for clustering and lensing that are both exact and  differentiable with respect to the parameters of the model. Conventional implementations of SHAM are based on iterative algorithms such as Richardson-Lucy deconvolution; here we use the JAX library for automatic differentiation to train \shamnet, a neural network that accurately approximates the stellar-to-halo mass relation (SMHM) defined by abundance matching. In our approach to  making differentiable predictions of large scale structure, we map parameterized PDFs onto each simulated halo, and calculate gradients of summary statistics of the galaxy distribution by using autodiff to propagate the gradients of the SMHM through the statistical estimators used to measure one- and two-point functions. Our techniques are quite general, and we conclude with an overview of how they can be applied in tandem with more complex, higher-dimensional models, creating the capability to make differentiable predictions for the multi-wavelength universe of galaxies.
\end{abstract}


\maketitle

\vspace{1cm}

\twocolumngrid



\section{Introduction}
\label{sec:intro}

Numerical simulations of cosmological structure formation can be viewed as prediction engines for the density field, and for the abundance and spatial distribution of dark matter halos. Of course, neither dark matter nor gravitationally self-bound halos are directly observed in the sky, and so additional modeling is required in order to transform a cosmological simulation into a prediction that is commensurable with astronomical measurements of galaxies. Contemporary simulations have by now achieved percent-level precision in the ability to characterize the density field, halo abundance and halo clustering in the nonlinear regime, whereas models connecting observed galaxies to the fundamental quantities in simulations lag far behind this level of accuracy \citep[see][for a recent review]{wechsler_tinker_2018}. Improving theoretical techniques for transforming cosmological simulations into predictions for the galaxy density field is thus a critical component of the precision-cosmology program.

In many of the structure formation models that are used for cosmological inference, predictions for the galaxy distribution are made by modeling galaxies as biased tracers of the underlying density field of matter. There are a wide range of techniques that can be used for this purpose, including the Zel'dovich approximation \citep{white_2014_zeldovich}, more general bias expansion methods based on either Lagrangian or Eulerian perturbation theory \citep{bernardeau_etal02,desjacques_etal18}, effective field theory \citep{carrasco_etal12}, and hybrid techniques that blend LPT with simulations \citep{kokron_etal21,derose_etal21b}. This approach to generating cosmological predictions has now been used by numerous galaxy surveys to derive constraints on the fundamental parameters of cosmology \citep[e.g.,][]{des_y3_3_by_2,joudaki_etal18}.

An alternative approach is to model the connection between dark matter halos and the galaxies residing within them. The conventional ``halo occupation model'' approach proceeds with an initial step in which a fitting function and/or a machine learning algorithm is calibrated to capture the cosmology-dependence of various summary statistics of dark matter halos, such as the halo mass function \citep[e.g.,][]{jenkins_etal01,mcclintock_etal19,bocquet_etal20} and halo bias \citep[e.g,][]{tinker_etal10,mcclintock_etal20}; once the approximations for these quantities are specified, models for the galaxy--halo connection determine the prediction for the clustering and lensing of galaxies, enabling the derivation of constraints on cosmological parameters \citep{cacciato_etal13,reddick_etal14,miyatake_etal21}.

As an alternative to traditional implementations of halo occupation modeling, it has become increasingly common to directly populate simulated halos with a Monte Carlo realization of the galaxy population; predictions for summary statistics of large scale structure are then made directly from the synthetic galaxy distribution using the same point estimators used to analyze observational data. In the conventional approach to halo occupation modeling, phenomena such as halo exclusion \citep{garcia_etal21_halo_exclusion}, satellite anisotropy \citep{sgro_etal13_satellite_anisotropy}, and galaxy assembly bias \citep{zentner_etal14} constitute a major technical challenge, particularly in the one-to-two-halo regime \citep{van_den_bosch_etal13}; one of the key advantages of the mock-population approach is that the numerical treatment of such effects is exact, with a precision limited only by the resolution and finite size of the simulation.

One of the principal sources of motivation for using mock-population techniques lies in the cosmological information content of the nonlinear regime. The potential to substantially improve cosmological constraints by incorporating smaller-scale information has been known for many years \citep{zentner_etal13,reid2014,krause_etal17}. Recent studies of the clustering of galaxies in the Baryon Oscillation Spectroscopic Survey (BOSS) have confirmed the long-forecasted constraining power of the nonlinear regime. In an analysis of the redshift-space clustering of the LOWZ galaxy sample \citep{lange_etal21}, the authors derived better-than 5\% constraints on the cosmological growth of structure, $f\sigma_8,$ a full factor of two stronger than any previous BOSS analysis that restricted attention to larger-scale measurements. Comparable gains in cosmological constraining power from the nonlinear regime of BOSS LOWZ galaxies were also found in \citet{wibking_etal20}. In closely related work analyzing the clustering of the Luminous Red Galaxy sample in eBOSS \citep{chapman_etal21}, it was found that constraints on $f\sigma_8$ are improved by 70\% when including information from nonlinear scales. 

Even when simulation-based predictions are made based on simple empirical models implemented in specialized libraries such as {\tt Halotools} \citep{hearin_etal17_halotools}, {\tt Corrfunc} \citep{corrfunc}, or AbacusHOD \citep{yuan_etal21_abacushod}, the computational demands of conducting Bayesian inference with survey-scale simulations are considerable. This challenge is tailor-made for a set of methods generally referred to as ``emulation'', which is a machine learning technique that is specifically designed for situations in which the behavior of some parametric function, $\mathcal{P}(\theta),$ is calculable, but expensive to evaluate. When confronted with such a situation, a natural way to proceed is to pre-compute $\mathcal{P}(\theta)$ for a finite collection of $\theta_{\rm i}$ that spans the domain of interest, and to use $\mathcal{P}(\theta_{\rm i})$ to train a machine learning algorithm to serve as an ``emulator'', or surrogate function, $\mathcal{F}(\theta),$ that approximates $\mathcal{P}(\theta).$ Once the emulator has been trained, one then proceeds to carry out performance-critical analyses such as MCMCs using $\mathcal{F}(\theta)$ rather than $\mathcal{P}(\theta),$ since the surrogate function is typically inexpensive to evaluate. Since the introduction of emulation methods to computational cosmology  \citep{heitmann_etal06}, these techniques have become a widely-used tool throughout the field \citep[e.g.,][]{kwan_etal15,nishimichi_etal19,euclid_emulator_pk_2020}.

The conventional approach to emulation is based on Gaussian Process (GP) regression. Briefly, in GP regression, it is assumed that the values $\mathcal{P}(\theta_{\rm i})$ are drawn from a multi-dimensional Gaussian distribution, $\mathcal{N}(\mathcal{P}(\theta_{\rm i}), \Sigma(\theta_{\rm i})),$ where $\Sigma(\theta_{\rm i})$ is a correlation matrix with hyper-parameters that are optimized during training \citep[see][for a contemporary review]{rasmussen_williams_GPbook_2006}. For example, in \citet{lawrence_etal17}, the abscissa $\theta_{\rm i}$ are cosmological parameters, and $\mathcal{P}(\theta_{\rm i})$ are values of the matter power spectrum; as another example, in \citet{zhai_etal19_aemulus3}, the abscissa reside in the joint space of cosmology and parameters of the Halo Occupation Distribution \citep[HOD,][]{berlind_weinberg02,zheng_eatl05}, and the ordinates are values of the redshift-space galaxy correlation function.

Although the GP approach to emulation has thus far been quite successful, this technique has limitations that create a significant impediment to applying it in tandem with the realistically complex models that will be required by near-future cosmological datasets. In \citet{zhai_etal19_aemulus3}, the authors found that the errors of their emulator were comparable to present-day measurement uncertainties on their predicted data vector, resulting in a nearly $50\%$ degradation of the cosmological constraints due to emulator noise. In principle, these inaccuracies could be remedied by increasing the number of training points $\theta_{\rm i};$ in practice, however, the size of the training data cannot be too large or the hyper-parameter optimization becomes computationally intractable due to the need to invert $\Sigma(\theta_{\rm i}).$ We point out that the effort in \citet{zhai_etal19_aemulus3} is one of the most ambitious GP applications in cosmology to date in the sense that the emulated parameter space includes parameters that jointly encode variations in cosmology as well as the galaxy--halo connection. And yet, the HOD model emulated in \citet{zhai_etal19_aemulus3} is essentially the simplest, lowest-dimensional model that can plausibly be used to interpret galaxy clustering data with present-day levels of uncertainty. Thus these shortcomings will only become more severe as observational cosmology progresses further into the 2020s: not only will the precision of cosmological measurements improve dramatically, but perhaps even more importantly, the dimension of the emulated model will need to increase substantially in order for the theoretical predictions to match the quality and richness of the data.

In the present work, we introduce a new theoretical framework for the galaxy--halo connection that is designed to address these issues. Our formalism differs from the typical simulation-based methodology outlined above in several respects. First, our simulation-based predictions for large scale structure observables are not based on stochastic Monte Carlo realizations; instead, we analytically propagate probability distributions from the galaxy--halo connection through to the corresponding summary statistics, and so the only source of stochasticity in our predictions derives from the finite size and resolution of the simulation. Second, instead of relying upon classical machine learning techniques such as Gaussian Process emulation, we instead rely on Artificial Intelligence (AI) algorithms for our surrogate functions. Since the neural networks implemented in contemporary deep learning libraries are routinely used in industry applications to approximate the behavior of million-parameter systems, our use of AI-based surrogate functions improves upon the problems associated with emulator accuracy and complexity outlined above. As a result of these techniques, our predictions for large scale structure observables are fully differentiable, end-to-end, enabling us to use gradient-based optimization and inference algorithms that exhibit much better scaling to the problem sizes that will characterize cosmological modeling in the 2020s.

In the present paper, for our model of the galaxy--halo connection, we use subhalo abundance matching \citep[SHAM,][]{kravtsov_etal04,conroy_etal06} as a toy model to demonstrate our framework; in \S\ref{sec:sham}, we provide a self-contained overview of SHAM. We train a neural network, \shamnet, that serves as a surrogate function approximating the SHAM mapping between stellar mass and halo mass; we describe our implementation of \shamnet in \S\ref{sec:shamnet}, relegating technical details to the appendices. Although any suitably formulated neural network is naturally differentiable, this is not the case for typical pipelines that make downstream predictions for the $n-$point functions of large scale structure; in \S\ref{sec:diffobs}, we describe our framework for constructing differentiable point estimators of the stellar mass function, galaxy-galaxy lensing, and galaxy clustering. 
We discuss our results in the broader context of related efforts in the literature in \S\ref{sec:discussion}, and we conclude by summarizing our primary findings in \S\ref{sec:conclusion}.

Throughout the paper, values of halo mass and distance are quoted assuming $h=1.$ For example, when writing $\mpeak=10^{12}\msun,$ we suppress the $\msun/h$ notation and write the units as $\msun.$ Values of stellar mass are quoted assuming $h=0.7.$

\section{Abundance Matching Primer}
\label{sec:sham}

The $\Mstar-\mhalo$ relation in traditional abundance matching is the unique, non-parametric mapping defined by equating the abundance of galaxies to the abundance of halos at the same redshift,
\beq
\label{eq:shamdef}
\Phi_{\rm g}(>\Mstar\vert z) = \Phi_{\rm h}(>\mhalo\vert z),
\eeq
where $\Phi_{\rm g}(>\Mstar)$ is the cumulative number density of galaxies, and $\Phi_{\rm h}(>\mhalo)$ is the cumulative number density of subhalos.
Contemporary forms of SHAM generalize Eq.\ref{eq:shamdef} to define non-parametric mappings between some subhalo property, $x_{\rm h},$ and some observed galaxy property, $y_{\rm g}.$ Application of the defining abundance matching equation guarantees that no matter the choice for  $x_{\rm h}$ and $y_{\rm g},$ the observed and predicted number density of galaxies will be in exact correspondence, by construction. We restrict the present investigation to $x_{\rm h}=\mpeak$ and $y_{\rm g}=\Mstar,$ and refer the reader to \S\ref{sec:discussion} for discussion of how our methodology could be extended to generalized galaxy/halo properties.

Even in the early literature on abundance matching \citep[e.g.,][]{tasitsiomi_etal04}, the importance of stochasticity in the $\Mstar-\mhalo$ relation was recognized to play an important role in the predictions of the model. Stochasticity in abundance matching is typically treated by assuming a model for the probability density $P(\Mstar\vert\mpeak)$ such as a log-normal, and then equating the number density of galaxies to the convolution of the subhalo mass function against the assumed PDF :
\beq
\label{eq:shamscatter}
\phi_{\rm g}(\Mstar) = \int_{0}^{\infty}dM_{\rm peak}\phi_{\rm h}(\mpeak)P(\Mstar\vert\mpeak),
\eeq
where $\phi_{\rm g}(\Mstar)$ is the differential number density of galaxies as a function of stellar mass, i.e. the stellar mass function (SMF), and $\phi_{\rm h}(\mpeak)$ is the differential number density of  subhalos as a function of mass, i.e. the subhalo mass function (SHMF).

In general, the $\Mstar-\mpeak$ relation defined by Equation \ref{eq:shamscatter} has no closed-form analytic solution, and so one must rely on approximations and numerical techniques to determine the scaling relation $\smhm$ that gives rise to the observed SMF when applied to the SHMF of a subhalo population. In \S\ref{subsec:deconvolution}, we discuss the conventional, non-parametric approach to determining the $\Mstar-\mpeak$ relation predicted by SHAM, and in \S\ref{subsec:shamparam} we describe the commonly-used alternative approach based on a parameterized approximation to SHAM, in both cases highlighting the close relationship between the SMF, the median relation $\smhm,$ and scatter in stellar mass at fixed halo mass.

\subsection{Non-parametric SHAM with scatter}
\label{subsec:deconvolution}

In non-parametric approaches to SHAM, the starting point is typically a catalog of simulated subhalos that defines the SHMF, and a volume-limited galaxy sample that defines the SMF. A particularly simple way to measure the cumulative abundance of galaxies or halos is simply to divide each object's rank-order by the volume containing the sample, although one may instead use a fitting function approximation to one or both abundance functions. However $\phi_{\rm g}$ and $\phi_{\rm h}$ are characterized, {\em in non-parametric SHAM the SMF and SHMF are held fixed,} and one numerically solves for the quantity $P(\Mstar\vert\mpeak)$ that is constrained by Eq.~\ref{eq:shamscatter}. Thus when generating a synthetic galaxy population with non-parametric SHAM, the stellar mass function of the mock is guaranteed to match the observed SMF, by construction, with an accuracy limited only by the convergence and robustness of the computational technique used to numerically solve  Eq.~\ref{eq:shamscatter}. In \S\ref{subsec:numerics} we describe two commonly used approaches to obtaining such a solution, and in \S\ref{subsec:shamsmhm} we highlight the basic features of the stellar-to-halo mass relation derived from the solution.

\subsubsection{Numerical Methods}
\label{subsec:numerics}

The most widely used numerical approach to solving Eq.~\ref{eq:shamscatter} is based on Richardson-Lucy deconvolution \citep{richardson72,lucy74}, an iterative algorithm  originally developed to recover a true underlying image that has been blurred by a point-spread function (PSF). In the context of abundance matching, we can think of the ``noisy image'' as $\phi_{\rm g}(\Mstar),$ i.e., the SMF that we measure in our observed galaxy sample; the PSF blurring the image is typically assumed to be a log-normal distribution in stellar mass at fixed halo mass; the RL deconvolution algorithm determines the ``true image'', $\phi^{\rm true}_{\rm g}(\Mstar),$ which in this case is the SMF that one {\em would} measure in the {\em absence} of any scatter in the $\Mstar-\mhalo$ relation. Once $\phi^{\rm true}_{\rm g}(\Mstar)$ is determined, then Eq.~\ref{eq:shamdef} is used to define $\smhm;$ together with the assumed level of log-normal scatter, this scaling relation then defines the quantity $P(\Mstar\vert\mpeak)$ that is used to map stellar mass onto a simulated subhalo population. In the conventional Monte Carlo based implementation, a synthetic galaxy population is generated by randomly drawing a value of $\Mstar$ from $P(\Mstar\vert\mpeak)$ for each subhalo; by construction, the resulting SMF of the synthetic population will agree with the SMF of the observed galaxy sample, $\phi_{\rm g}(\Mstar),$ again with accuracy limited only by the RL deconvoluation-based estimation of $\phi^{\rm true}_{\rm g}(\Mstar)$ \citep[see][Section 3, for further details]{behroozi_etal10}.

An alternative approach presented in \citet{kravtsov_etal18} is instead based on the assumption that the $\Mstar-\mpeak$ relation can be {\em locally} approximated as a power law. Under this assumption, for any particular value of $\mpeak$ and level of scatter, it is straightforward to solve for the power-law normalization and index that produces the observed $\phi_{\rm g}(\Mstar)$ from the simulated $\phi_{\rm h}(\mpeak).$ By repeating this exercise at a finite set of control points in $\mpeak$ that densely spans the relevant range, one can use the results of the computation as an interpolation table that defines the $\smhm$ scaling relation \citep[see][Appendix A, for further details]{kravtsov_etal18}.

\subsubsection{Stellar-to-Halo Mass Relation of Non-Parameteric SHAM}
\label{subsec:shamsmhm}

\begin{figure}
\includegraphics[width=8cm]{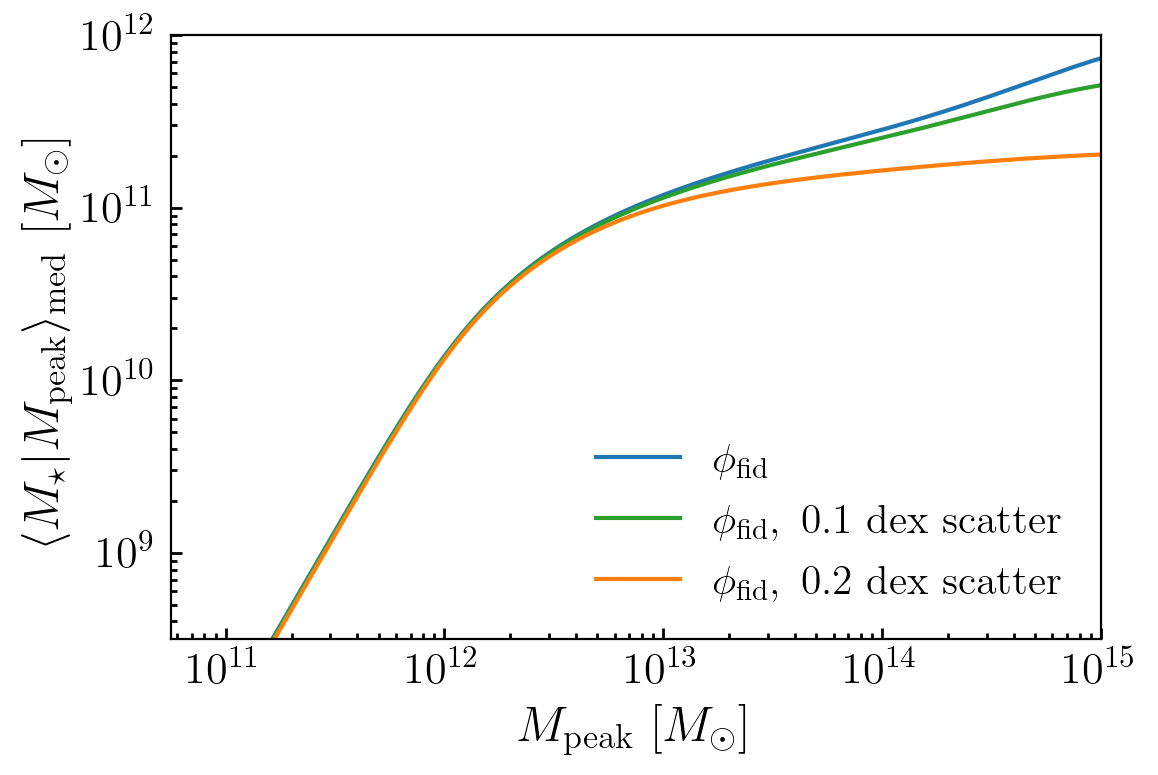}
\caption{{\bf Role of scatter in non-parametric SHAM}. Each curve shows the $\Mstar-\mhalo$ relation of a SHAM model that predicts the same stellar mass function, $\phi_{\rm fid}(\Mstar),$ but with different levels of scatter in $\Mstar$ at fixed halo mass, as indicated in the legend. For each model, the median relation $\smhm$ is plotted on the vertical axis as a function of halo mass, $\mpeak.$}
\label{fig:shamscat}
\end{figure}

Figure \ref{fig:shamscat} gives a simple demonstration of the role of scatter in non-parametric abundance matching. Each curve shows the abundance matching prediction for $\smhm$ using the same stellar mass function, and the same subhalo mass function, but with different levels of scatter as indicated in the legend. For our SMF, $\phi_{\rm fid},$ we use a Schechter function with best-fit parameters taken from \citet{panter_etal07}, and for the subhalo mass function we use the fitting function presented in Appendix \ref{sec:shmf}. We calculate $\smhm$ using a publicly available python wrapper\footnote{\url{https://bitbucket.org/yymao/abundancematching/src}} of the RL deconvolution implementation originally developed in \citet{behroozi_etal10}.

All curves in Figure \ref{fig:shamscat} present the same qualitative shape of a double power-law, with stronger levels of scatter producing shallower power-law relations at the high-mass end. We can understand this characteristic flattening in terms of Eddington bias \citep{Eddington13}: stochasticity in the $\Mstar-\mpeak$ relation will naturally result in proportionally more up-scatter above a threshold relative to down-scatter below the threshold, simply because low-mass halos are more abundant than high-mass halos. Thus it is sensible that we see stronger effects of scatter at high mass, where the slope of the mass function is rapidly steepening. We return to this issue in the subsequent section on parametric approximations to abundance matching.

\begin{figure}
\includegraphics[width=8cm]{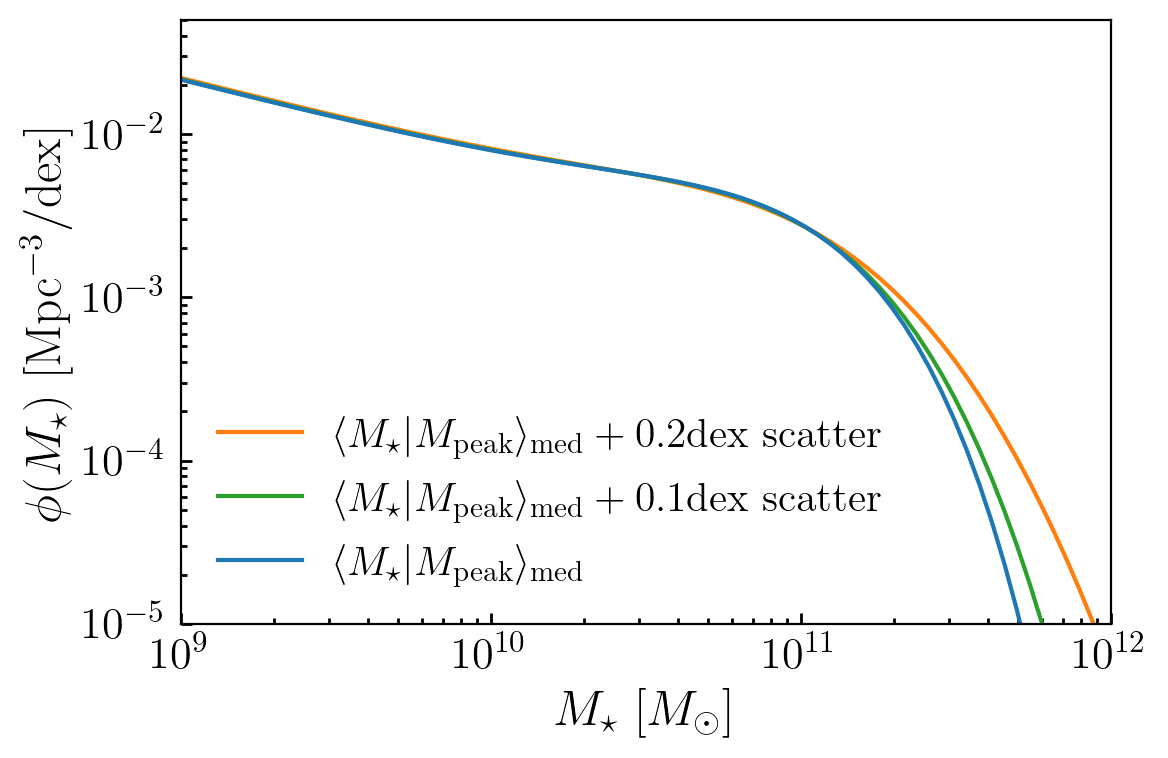}
\caption{{\bf Role of scatter in parameterized abundance matching}. Each curve shows the stellar mass function predicted by a parameterized abundance matching model with the same $\smhm,$ but with different levels of scatter in $\Mstar$ at fixed halo mass. For each model, the number density of galaxies, $\phi(\Mstar),$ is plotted on the vertical axis as a function of stellar mass.}
\label{fig:smhmscat}
\end{figure}

\subsection{Parametric Approximations to SHAM with scatter}
\label{subsec:shamparam}

In studies of the galaxy-halo connection connecting stellar mass to halo mass, numerous models have directly parameterized the scaling relation $\smhm,$ rather than using numerical methods of solution to Eq.~\ref{eq:shamscatter} \citep[e.g.,][]{moster_etal10, moster_etal13, behroozi_etal13,rodriguez_puebla_etal15_smhm_colors}. In such models, the assumed functional forms for $\smhm$ all have the same general double power-law shape shown in Figure \ref{fig:shamscat}; the free parameters of these models regulate the low- and high-mass slopes, as well as the normalization and shape of the transition between the two regimes.

It has been shown explicitly that the $\smhm$ predicted via RL deconvolution can indeed be accurately approximated with the functional forms used in \citet{behroozi_etal10} and related studies, and so it is common for such parametric models to be informally referred to as ``abundance matching.'' However, we point out a clear distinction in the way these two models utilize information from observations: in parameterized models of $\smhm,$ there is no guarantee that the observed stellar mass function will be correctly predicted, as is the case with non-parametric abundance matching. Instead, in parameterized models, observations of the SMF are treated as constraining data that can be used to derive confidence intervals on the parameters controlling the stellar-to-halo mass relation. Thus parametric vs. non-parametric SHAM models use information from observations in quite a different manner when evaluating the likelihood of a proposed point in parameter space.

Figure \ref{fig:smhmscat} illustrates the role of scatter in predictions for the stellar mass function made by parameterized abundance matching. The axes show the number density of galaxies as a function of stellar mass, $\phi(\Mstar).$ Each curve shows results for models with different levels of scatter about the same $\Mstar-\mhalo$ relation. Again we see the proportionally larger influence of Eddington bias at higher mass, due to up-scatter from lower-mass halos that outnumber higher-mass halos.

\section{SHAMNet}
\label{sec:shamnet}

In this section, we describe a new approach to abundance matching based on \shamnet: a neural network approximation to the numerical solution to Eq.~\ref{eq:shamscatter}. We begin in \S\ref{subsec:shamnet_overview} with a high-level overview of how \shamnet is defined. In \S\ref{subsec:shamnet_ingredients} we describe the key ingredients we use in our implementation, and in \S\ref{subsec:shamnet_definition} we describe how we have used these ingredients to train \shamnet to approximate the abundance matching prediction for the SMHM scaling relation, $\smhm.$  We give a detailed account of the architecture and training of \shamnet in the appendices; our source code is publicly available at \url{https://github.com/ArgonneCPAC/shamnet}, and is available for installation with pip.

\subsection{SHAMNet Overview}
\label{subsec:shamnet_overview}

As discussed in \S\ref{sec:sham}, the SHAM technique defines the probability distribution $P(\Mstar\vert\mpeak)$ such that when this PDF is convolved against the subhalo mass function, $\phi_{\rm h}(\mpeak),$ the galaxy stellar mass function, $\phi_{\rm g}(\Mstar),$ is recovered. Under the assumption of log-normal scatter, this PDF is fully described by its first and second moments, $\smhm$ and $\sigma(\mpeak),$ respectively. Thus the goal of any SHAM implementation is to accept ingredients for $\phi_{\rm g}(\Mstar),$ $\phi_{\rm h}(\mpeak),$ and $\sigma(\mpeak)$ as inputs, and to return the scaling relation $\smhm$ that provides a solution to Eq.~\ref{eq:shamscatter}. \shamnet is simply a neural network that provides a mapping from these ingredients to the desired stellar-to-halo mass relation.

In building \shamnet, we assume that the observed SMF can be characterized with sufficient precision using some parametrized functional form, $\phi_{\rm g}(\Mstar\vert\theta_1);$ we similarly assume that the subhalo mass function can be parametrically described as $\phi_{\rm h}(\mpeak\vert\theta_2);$ finally, we assume log-normal scatter in stellar mass at fixed halo mass, allowing the level of scatter, $\sigma,$ to be a parametrized function of halo mass, $\sigma(\mpeak\vert\theta_3).$ For notational convenience, we will use the generic variable $\theta$ to refer to the collection of these parameters, so that $\theta$ fully specifies $\phi_{\rm g}(\Mstar),$ $\phi_{\rm h}(\mpeak),$ and $\sigma(\mpeak).$ \shamnet is defined to be a neural network that accepts $\theta$ and $\mpeak$ as input, and returns $\Mstar$ as output; the network parameters of \shamnet are trained so that Eq.~\ref{eq:shamscatter} is satisfied by the mapping.

\subsection{SHAMNet ingredients}
\label{subsec:shamnet_ingredients}
Our formulation of \shamnet requires a parametric description of $\phi_{\rm g}(\Mstar),$ $\phi_{\rm h}(\mpeak),$ and $\sigma(\mpeak).$ In this section, we describe our models for each of these ingredients in turn.

Our parameterization of $\phi_{\rm g},$ the differential number density of galaxies, is based on the Schechter function,  $\phi_{\rm S},$ defined as
\beq
\label{eq:schechter}
\\
\phi_{\rm S}(x\vert x_{\ast}, \phi_{\ast}, \alpha) = {\rm ln}(10)\phi_{\ast}10^{(x-x_{\ast})\cdot(\alpha+1)}\exp(-10^{x-x_{\ast}})\nonumber,
\eeq
where $x=\log_{10}\Mstar$ in units of $\msun.$ Our galaxy SMF is thus characterized by $x_{\ast}, \phi_{\ast},$ and $\alpha,$ and we define \shamnet with a fixed value $x_{\ast}=10.85,$ closely mimicking the SMF in the low-redshift universe \citep{li_white_2009}.

We parameterize the subhalo mass function in terms of the cumulative number density of subhalos, $\Phi_{\rm h}(>\mpeak\vert z),$ using the fitting function described in Appendix \ref{sec:shmf}. Briefly, $\Phi_{\rm h}(>\mpeak\vert z)$ behaves like a power-law at low mass, with a normalization $A_{\rm h},$ and an index $\beta_{\rm h},$ with an exponential cutoff the high-mass end characterized by a cutoff mass, $x_{\rm h},$ and cutoff speed, $k_{\rm h}.$ In defining \shamnet, we hold the subhalo mass function parameters fixed to the values supplied in Appendix \ref{sec:shmf} that have been tuned to closely match the SHMF in the Bolshoi-Planck simulation \citep[BPL,][]{klypin_etal11}.

We model the scatter in $\Mstar$ at fixed $\mpeak$ as a log-normal distribution with a width that is allowed to vary as a function of halo mass, $\sigma(\mpeak).$ For the mass-dependence of the scatter, we use the sigmoid function given in \ref{eq:sigmoid}, with the parameters $y^{\sigma}_{\rm lo}$ and $y^{\sigma}_{\rm hi}$ specifying the level of scatter at the low- and high-mass end, respectively, and with $k^{\sigma}=1$ and $x^{\sigma}_0=12$ held fixed:
\beq
\sigma(x\vert y_{\rm lo}^{\sigma},y_{\rm hi}^{\sigma}) = \mathcal{S}(x, 12, 1, y_{\rm lo}^{\sigma},y_{\rm hi}^{\sigma}),
\eeq where $x=\log_{10}\mpeak.$

\subsection{SHAMNet Definition}
\label{subsec:shamnet_definition}

With the ingredients defined in \S\ref{subsec:shamnet_ingredients}, the parameters $\theta$ collectively describe our model for the SMF, $\phi_{\rm g}(\Mstar\vert\theta),$ the subhalo mass function, $\phi_{\rm h}(\mpeak\vert\theta),$ and the level of log-normal scatter, $\sigma(\mpeak\vert\theta).$ For any particular value of $\theta,$ we seek to identify the stellar-to-halo mass relation, $\smhmtheta,$ that provides the self-consistent relationship between these quantities (as defined by Eq.~\ref{eq:shamscatter}). We define \shamnet, $\fsham,$ to be a neural network that has been trained to supply the appropriate scaling relation:
\beq
\label{eq:shamnetdef}
\fsham(\mpeak,\theta\vert\psi)\equiv\smhmtheta,
\eeq
where the variables $\psi$ are the weights and biases of the neural network. Thus for any particular values of $\psi,$ the \shamnet function $\fsham$  accepts $\mpeak$ and $\theta$ as input, and returns $\Mstar$ as output.

The objective of training \shamnet is to optimize the weights and biases $\psi$ such that the resulting mapping in Eq.~\ref{eq:shamnetdef} supplies a stellar-to-halo mass relation that satisfies Eq.~\ref{eq:shamscatter} to the required precision. Once trained, the parameters $\psi$ are thereafter held fixed, and $\fsham(\mpeak,\theta)=\Mstar$ defines the stellar-to-halo mass relation for any input combination of SMF, subhalo mass function, and scatter. 

\begin{figure}
\includegraphics[width=8cm]{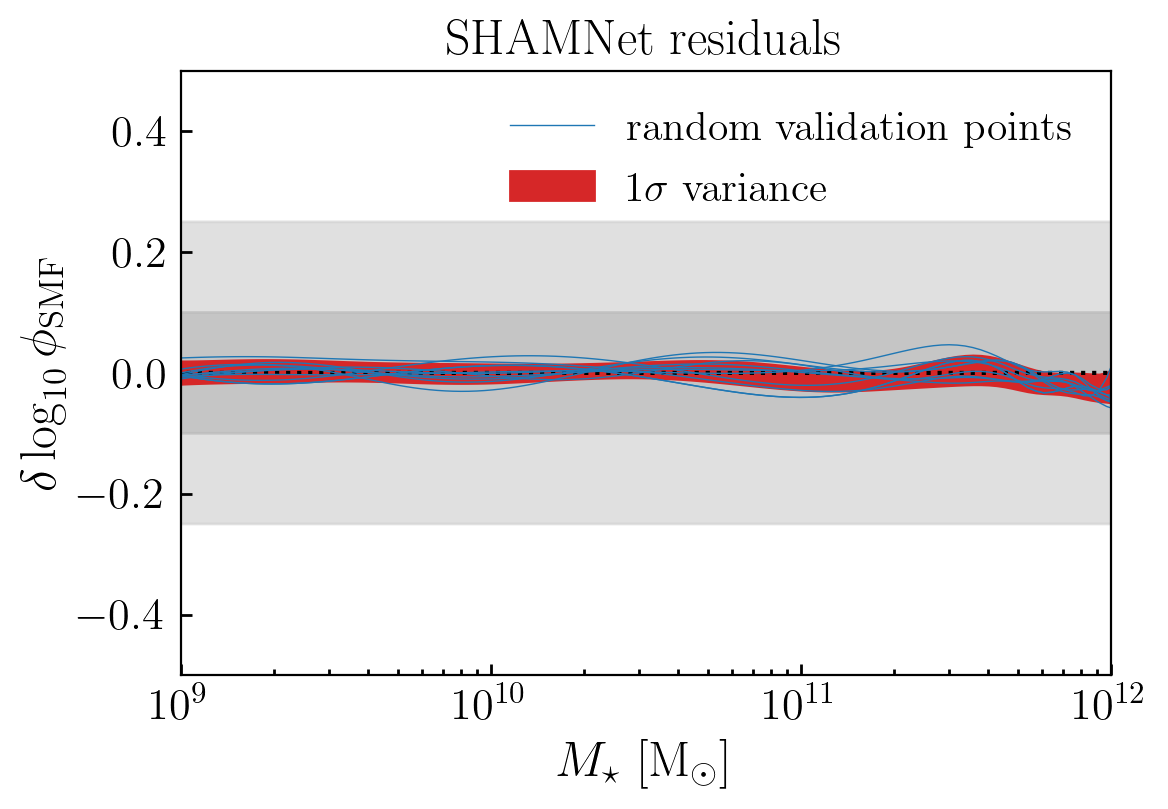}
\caption{{\bf Accuracy of \shamnet}. The vertical axis shows the logarithmic difference between the target stellar mass function and the SMF that results from the \shamnet prediction for the stellar-to-halo-mass relation. The red band shows the variance in the residual error amongst 100 randomly selected points in the SMF parameter space; the blue curves show the particular residuals for 10 of these points.}
\label{fig:shamnet_accuracy}
\end{figure}

As shown in Figure~\ref{fig:shamnet_accuracy}, accurately training \shamnet does not present a significant challenge for a simply-connected network with only a few short layers. On the vertical axis, we show the logarithmic difference between target SMF defined by $\theta,$ and the SMF that results from the \shamnet prediction for $\smhmtheta.$ The level of success shown in Fig.~\ref{fig:shamnet_accuracy} should not be surprising, since neural networks are routinely called upon to approximate far more complex functions than abundance matching. We refer the reader to Appendix~\ref{sec:shamnet_train} for detailed information on how we trained \shamnet.

Beyond its accuracy and convenience, the function $\fsham(\mpeak,\theta)$ is an analytically differentiable function of the parameters $\theta,$ because the behavior of the function is simply the composition of a chain of $C^{\infty}$ functions. As described in Appendix~\ref{sec:shamnet_train}, our implementation of \shamnet in the JAX library makes it straightforward and efficient to compute these gradients via automatic differentiation. In \S\ref{sec:diffobs} below, we describe how these \shamnet gradients can be propagated to permit calculation of derivatives of large scale structure observables such as two-point clustering and lensing.

\section{Differentiable Predictions for Large Scale Structure}
\label{sec:diffobs}

In \S\ref{sec:shamnet}, we described how \shamnet, $\fsham,$ serves as a surrogate function that approximates the stellar-to-halo mass relation of abundance matching, $\fsham(\mpeak\vert\theta)\equiv\smhmtheta.$ We remind the reader that the parameters $\theta$ control the behavior of the galaxy stellar mass function, $\phi_{\rm g}(\Mstar\vert\theta),$ the subhalo mass function, $\phi_{\rm h}(\mpeak\vert\theta),$ and halo mass-dependent scatter, $\sigma(\mpeak\vert\theta).$ For the remainder of the paper, our principal focus will be on leveraging the {\em differentiability} of $\fsham(\mpeak\vert\theta)$ with respect to the parameters $\theta.$ Physically, the gradients $\partial\fsham/\partial\theta$ encode how the stellar-to-halo mass relation changes in response to changes in $\phi_{\rm g}(\Mstar),$ $\phi_{\rm h}(\mpeak),$ and $\sigma(\mpeak).$

In conventional SHAM implementations, the scaling relation $\smhmtheta$ is computed numerically via an iterative algorithm such as RL deconvolution, and so one must rely on finite differencing methods to estimate the gradients of the scaling relation with respect to the parameters $\theta.$ By contrast, our JAX implementation of $\fsham$ allows us to calculate $\partial\smhmtheta/\partial\theta$ to machine precision with high efficiency. In this section, we describe a set of techniques that will allow us to leverage the availability of these gradients to directly differentiate simulation-based forward-modeling predictions for large scale structure observables. 

\subsection{Stellar Mass Function}
\label{subsec:diffsmf}

Simulation-based predictions for large scale structure observables are typically based on Monte Carlo realizations of a model applied to a (sub)halo catalog. Using such Monte Carlo methods, the SHAM prediction for the stellar mass function proceeds in two steps:
\ben
\item Assign a value $\Mstar$ to every subhalo in the catalog by randomly drawing from a log-normal PDF centered at $\smhm$ with scatter $\sigma.$
\item For the $i^{\rm th}$ bin of the SMF, $\phi_{\rm g}(\Mstar^{\rm i}),$ sum the number of synthetic galaxies with $\Mstar^{\rm i}<=\Mstar<\Mstar^{\rm i+1},$ and divide by the normalization factor appropriate for the population volume and bin size, $V_{\rm i}.$
\een
In order to calculate gradients of the predicted SMF with respect to model parameters $\theta,$ the above procedure is repeated for parameter values $\theta$ that have been perturbed about some fiducial value, and the gradient is estimated via finite-differencing methods, e.g., $\Delta\phi_{\rm g}(\Mstar)/\Delta\theta.$

Here we consider an alternative approach to predicting $\phi_{\rm g}(\Mstar)$ that does not rely on a Monte Carlo realization. For the $\alpha^{\rm th}$ subhalo in the catalog with halo mass $\mpeak^{\alpha},$ we can analytically calculate the probability that the galaxy mapped onto the subhalo will fall within the ${\rm i^{th}}$ bin:
\beq
\label{eq:diffweight}
w_{\rm i}^{\alpha} &=& \int_{\Mstar^{\rm i}}^{\Mstar^{\rm i+1}}{\rm d\Mstar}P(\Mstar\vert\mpeak^{\alpha}).
\label{eq:diffsmf}
\eeq
Once all the weights $w_{\rm i}^{\alpha}$ have been calculated for each subhalo in the catalog, then predicting the SMF is simply a matter of summing the weights in each bin and normalizing the result:
\beq
\label{eq:diffsmf}
\phi_{\rm g}(\Mstar^{\rm i}) &=& \sum_{\alpha}w_{\rm i}^{\alpha}/V_{\rm i}.
\eeq

\begin{figure}
\centering
\includegraphics[width=8cm]{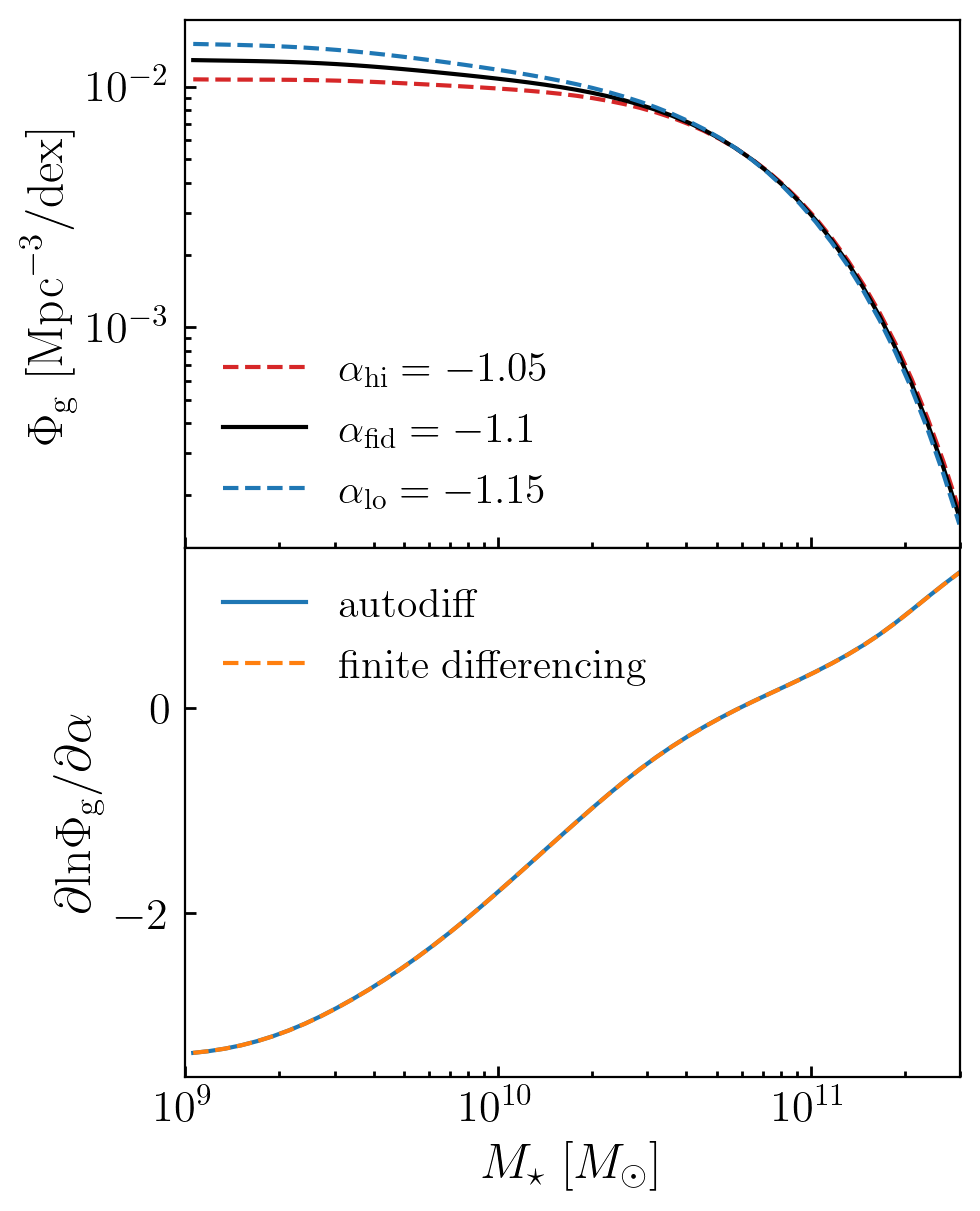}
\caption{{\bf Gradients of simulation-based SMF predictions}. In the top panel we show three different SMFs, with different values of $\alpha$ as indicated in the legend. In the bottom panel we show the logarithmic derivative of the SMF with respect to $\alpha,$ calculating the result with both finite-differencing methods and automatic differentiation.}
\label{fig:smf_grads}
\end{figure}

Operationally, the only difference between these two approaches is that we compute the integrals in Eq.~\ref{eq:diffweight} analytically rather than via Monte Carlo integration. While Monte Carlo-based predictions will be statistically commensurable with the SMF computed via Eq.~\ref{eq:diffsmf}, the analytical approach can be more computationally efficient because numerical convergence may require a large number of realizations, particularly at the high-mass end due to the rarity of massive galaxies. Beyond this computational advantage, from Equations \ref{eq:diffweight}-\ref{eq:diffsmf} we can readily see that the gradient $\partial\phi_{\rm g}(\Mstar)/\partial\theta$ can be calculated exactly, since it is simply a sum of $C^{\infty}$ functions:
\beq
\label{eq:smf_grads}
\frac{\partial}{\partial\theta}\phi_{\rm g}(\Mstar^{\rm i}) = \frac{1}{V_{\rm i}}\sum_{\alpha}\frac{\partial w_{\rm i}^{\alpha}}{\partial\theta}
\eeq

In Figure~\ref{fig:smf_grads}, we show how the stellar mass function responds to changes with respect to $\alpha,$ the parameter controlling the power-law slope of the Schechter function defined in Eq.~\ref{eq:schechter}. In the top panel we show three different SMFs, with different values of $\alpha$ as indicated in the legend. To compute $\phi_{\rm g}(\Mstar),$ we used Eq.~\ref{eq:diffsmf} to calculate the weight attached to every subhalo of the BPL simulation at $z=0,$ using 50 logarithmically-spaced bins spanning $10^9\msun<\Mstar<10^{11.5}\msun.$ In the bottom panel of Figure~\ref{fig:smf_grads}, we show the logarithmic derivative of the SMF with respect to $\alpha.$ For the curve labeled ``finite differencing", we calculated the gradient numerically by repeatedly calculating the result of Eq.~\ref{eq:diffsmf} for low and high values of $\alpha.$ For the curve labeled ``autodiff", we use JAX to calculate $\partial w_{\alpha}/\partial\theta$ for every subhalo, and then we use numba to propagate these derivatives through to the computation of the SMF according to Eq.~\ref{eq:smf_grads}.

\subsection{Galaxy Lensing}
\label{subsec:diffds}

In this section, we adapt the methods described in \S\ref{subsec:diffsmf} to make differentiable predictions for galaxy lensing, $\dsrp,$ the excess surface mass density at a projected distance $R$ from the center of a stacked sample of galaxies, defined as
\beq
\label{eq:dsdef}
\dsrp\equiv\bar{\Sigma}(<R) - \Sigma(R).
\eeq
In Eq.~\ref{eq:dsdef}, $\Sigma(R)$ is the surface mass density projected along the line-of-sight to the stack, and $\bar{\Sigma}(<R)$ is the value of $\Sigma$ averaged over the cylinder interior to $R.$

Using the Halotools implementation of the technique presented in \citet{lange_etal19} and reviewed in Appendix \ref{sec:simlensing}, the profiles $\dsrp$ can be computed in advance on a per-object basis for every subhalo in a simulated snapshot. Once tabulated, the lensing signal produced by a stack of galaxies can be computed as the average signal produced by each of the subhalos in the sample,
\beq
\label{eq:dsfast}
\langle\dsrp\rangle = \langle\ds^{\alpha}(R)\rangle_{\alpha\in\ {\rm sample}},
\eeq
where $\ds^{\alpha}$ is the lensing of the $\alpha^{\rm th}$ subhalo in the catalog.

Traditional SHAM predictions for galaxy lensing are straightforward to compute using standard Monte Carlo methods. First, a value of $\Mstar^{\alpha}$ is mapped onto every subhalo in the snapshot according to a random draw from the appropriate log-normal. The SHAM prediction for the lensing signal produced by a stack of galaxies is then given simply by Eq.~\ref{eq:dsfast}, taking the average over only those galaxies with randomly drawn values of $\Mstar^{\alpha}$ that satisfy the desired selection criteria, e.g., $\Mstarlo<\Mstar^{\alpha}<\Mstarhi.$

\begin{figure}
\centering
\includegraphics[width=8cm]{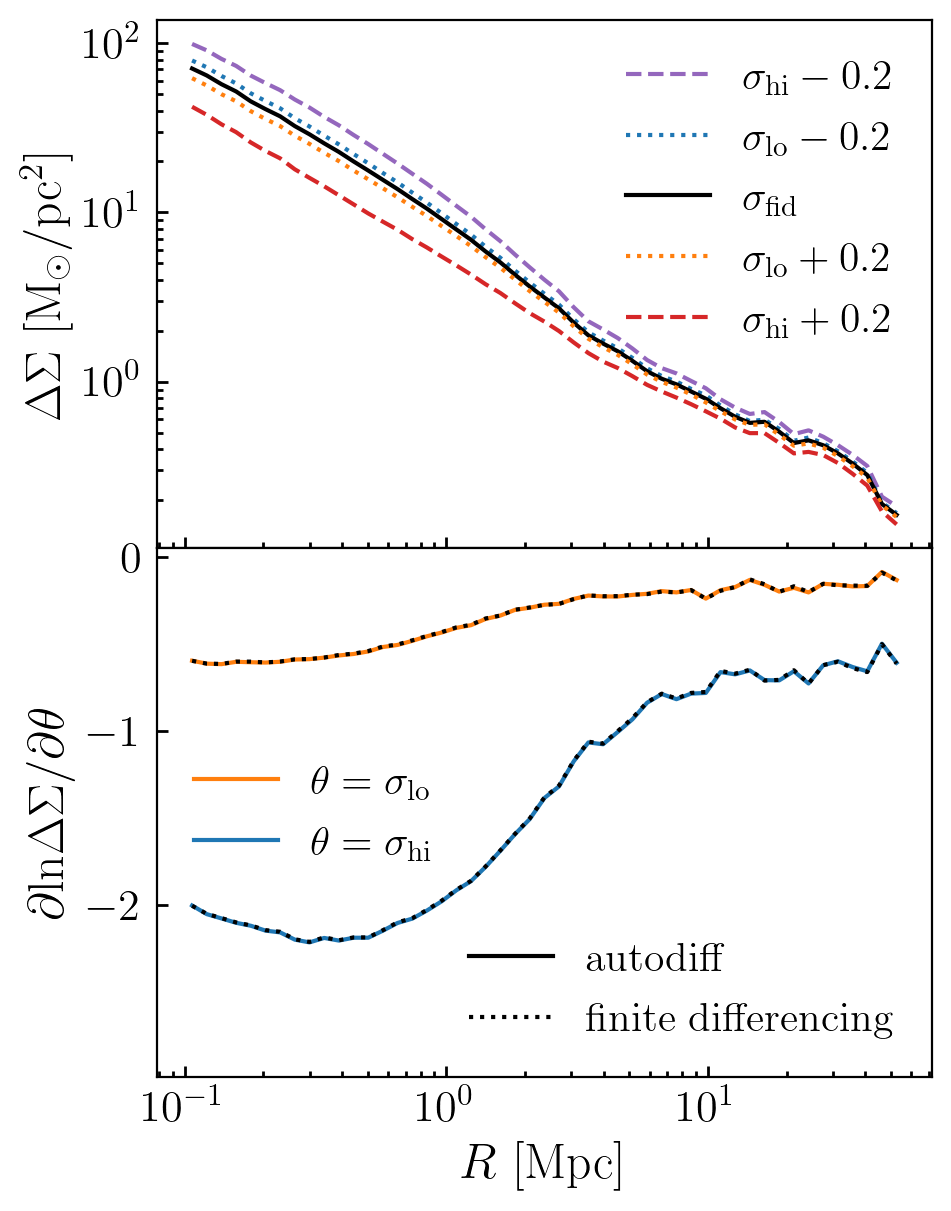}
\caption{{\bf Gradients of simulation-based $\Delta\Sigma$ predictions}. In the top panel, we show $\Delta\Sigma$ for galaxies with $\Mstar>10^{11}\msun$ residing in subhalos in the BPL simulation, with different curves corresponding to values of the parameters controlling mass-dependent scatter, as indicated in the legend. In the bottom panel we show the logarithmic derivative of $\Delta\Sigma,$ calculating the result with both finite-differencing methods and automatic differentiation.}
\label{fig:ds_grads}
\end{figure}

As a differentiable alternative to Monte Carlo-based predictions, we can instead compute the left-hand side of Eq.~\ref{eq:dsfast} as a {\em weighted} average, where the weight attached to each subhalo, $w_{\alpha},$ is the probability that the galaxy passes the selection function, calculated according to Eq.~\ref{eq:diffweight}:
\beq
\label{eq:diffds}
\langle\dsrp\rangle = \frac{\sum_{\alpha}w_{\alpha}\cdot\ds^{\alpha}(R)}{\sum_{\alpha}w_{\alpha}},
\eeq
where the summations in Eq.~\ref{eq:diffds} are performed over all subhalos with non-zero probability of residing in the stellar mass bin. In the limit of an infinite simulated volume, Eq.~\ref{eq:dsfast} converges to Eq.~\ref{eq:diffds}, but the latter method has the advantage of being differentiable with respect to parameters $\theta$ of the galaxy--halo connection:
\beq
\label{eq:dsgradient}
\frac{\partial}{\partial\theta}\langle\dsrp\rangle = \sum_{\alpha}\frac{\partial}{\partial\theta}\left(\frac{w_{\alpha}}{\sum_{\alpha}w_{\alpha}}\right)\ds^{\alpha}(R).
\eeq
To compute the left-hand side of Eq.~\ref{eq:dsgradient}, we use JAX to calculate the quantity $\frac{\partial}{\partial\theta}(w_{\alpha}/\sum_{\alpha}w_{\alpha})$ for each simulated subhalo, and we then we use the python library numba \citep{lam2015numba} to perform the weighted-sum on the right-hand-side of Eq.~\ref{eq:dsgradient} using the precomputed lensing profile of each individual subhalo. 

In the top panel of Figure \ref{fig:ds_grads}, we show a few examples of computations of $\Delta\Sigma$ for galaxies with $\Mstar>10^{11}\msun$ based on the BPL simulation, with different curves corresponding to different values of the parameters controlling mass-dependent scatter. For the fiducial model shown with the black curve, we use $(\sigma_{\rm lo},\sigma_{\rm hi})=(0.4, 0.25);$ results based on independent perturbations to $\sigma_{\rm lo}$ and $\sigma_{\rm hi}$ are color-coded as indicated in the legend. In the bottom panel of Figure~\ref{fig:ds_grads}, we show gradients of $\Delta\Sigma$ with respect to $\sigma_{\rm lo}$ and $\sigma_{\rm hi},$ again comparing the results of the computation based on autodiff vs. finite-differencing methods. Due to the power-law shape of the subhalo mass function, larger scatter in the SMHM at {\it any} mass corresponds to a lower amplitude of $\Delta\Sigma,$ since increasing the scatter results in a higher proportion of low-mass, weakly-clustered subhalos that up-scatter into the sample. The effect on $\Delta\Sigma$ for this relatively massive galaxy sample is more pronounced for $\sigma_{\rm hi};$ this is sensible, since at high mass the slope of the mass function is falling off rapidly, and the halo bias function is rapidly steepening.

\subsection{Galaxy Clustering}
\label{subsec:difftpcf}

The computation of differentiable two-point functions proceeds in much the same way as that for galaxy lensing. Each object again has a probability $w_{\alpha}$ of being in the sample. The product of the weights, $w_{\alpha}\cdot w_{\beta}$, is interpreted as the probability that a given pair of objects would be counted when accumulating the number of pairs at a given separation for, e.g., computing a two-point function. Formally, we are assuming that when conditioned on the values of the weights themselves, the presence of any given object in the sample is independent of all of the others. Finally, there exists some ambiguity in how exactly to define an unclustered or ``random" sample given a set of objects with weights. We define the random sample as a set of points with a random spatial distribution, each with a weight that has been randomly assigned from the original sample. We show below that this definition properly reproduces zero clustering signal for samples where points are distributed randomly, and only kept with a probability equal to the weight.

With these assumptions, we use the following estimator for the two-point correlation function
\beq
\hat{\xi}_{i} = \frac{DD_{i}}{RR_{i}} - 1
\eeq
where
\beq
DD_{i} = \sum_\alpha\sum_{\beta\neq\alpha} w_{\alpha} w_{\beta} {\cal B}_{i}(x_{\alpha}, x_{\beta})
\eeq
and
\beq
RR_{i} = \left(\sum_{\alpha}w_\alpha\right)^{2}\left(1 - \frac{1}{N_{\rm eff}}\right) \frac{V_i}{V}\ .
\eeq
Here ${\cal B}_{i}(x_{\alpha}, x_{\beta})$ is the bin selection function, which is unity if points $\alpha$ and $\beta$ are separated such that their separation falls in bin $i$, and zero otherwise; $V_i$ is the volume of the bin, and $V$ is the total volume of the domain; finally, the quantity $N_{\rm eff}$ is defined as:
\beq
N_{\rm eff} \equiv \frac{\left(\sum_\alpha w_\alpha\right)^2}{\sum_\alpha w_\alpha^2},
\eeq
so that $N_{\rm eff}$ is interpreted as the \textit{effective sample size}.

To better understand the role of the denominator, $RR_i,$ we can rewrite the estimator as
\beq
RR_i = \left(\sum_\alpha w_\alpha \left(\sum_\beta w_\beta - w_\alpha\right) \right) \times \frac{V_i}{V}
\eeq
In this form, we can see that this estimator directly computes the expected pair counts, excluding self-pairs, if we have randomly assigned the weights to the random sample. This estimator is equivalent to the $\mathcal{O}(\frac{1}{N})$ corrections for finite samples described in \href{https://nbviewer.org/gist/lgarrison/1efabe4430429996733a9d29397423d2}{the Corrfunc documentation} \citep{corrfunc}, but with weights that represent membership probabilities. Finally, the factor $V_i/V$ is the fraction of the total volume occupied by the $i^{\rm th}$ bin. 
\begin{figure}
\centering
\includegraphics[width=8cm]{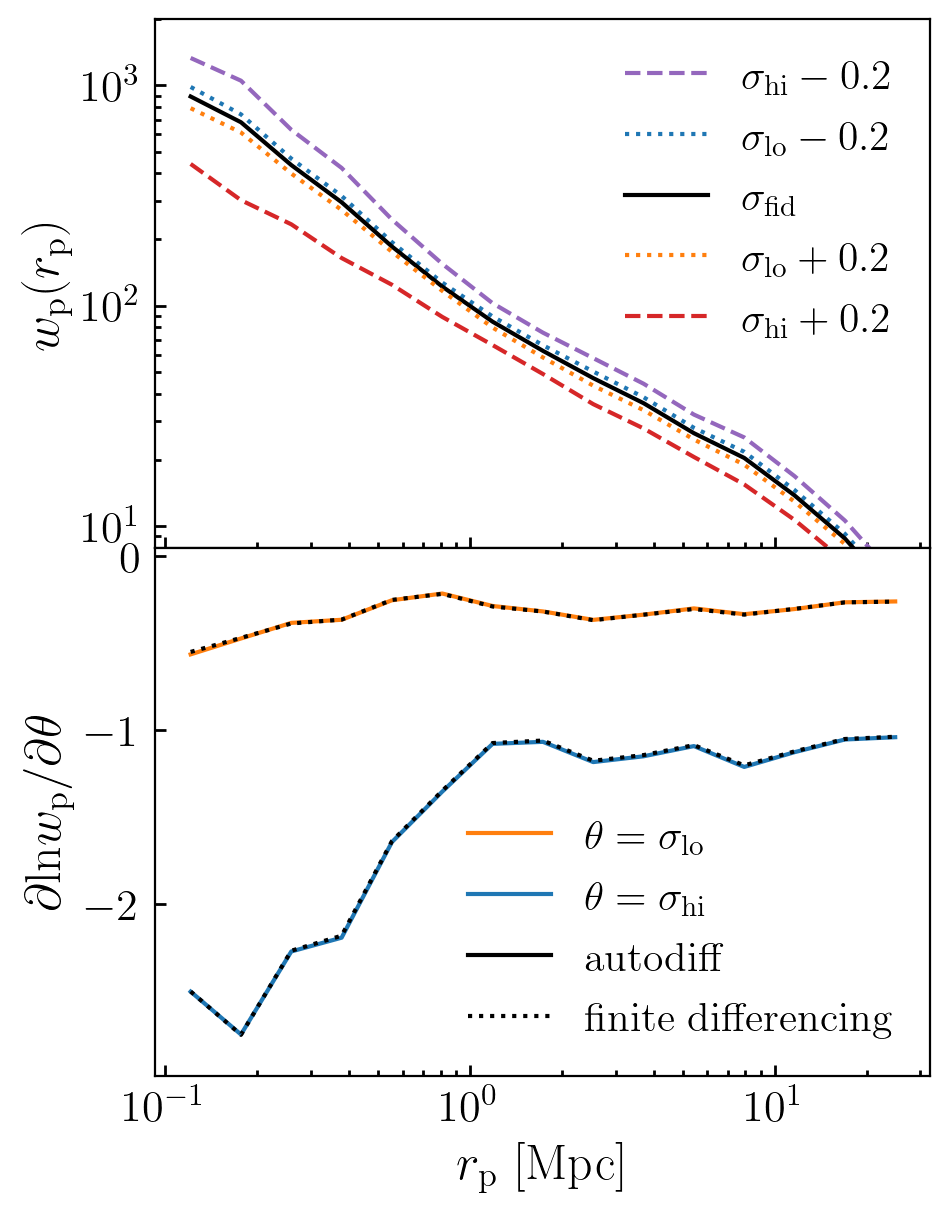}
\caption{
{\bf Gradients of simulation-based $\wproj$ predictions.} Same as Fig.~\ref{fig:ds_grads}, but for $\wprp.$ In the top panel, different curves show $\wprp$ for different values of the parameters controlling mass-dependent scatter, as indicated in the legend. In the bottom panel we show the logarithmic derivative of $\wprp,$ calculating the result with both finite-differencing methods and automatic differentiation.}
\label{fig:wp_grads}
\end{figure}

From this definition of the two-point clustering estimator, we can use the chain rule to compute the derivative,
\beq
\frac{\partial \hat{\xi}_{i}}{\partial \theta} = \frac{1}{RR_i}\frac{\partial DD_i}{\partial \theta} - \frac{DD_{i}}{RR_{i}^2}\frac{\partial RR_i}{\partial \theta}
\eeq
where
\beq
\\
\frac{\partial DD_i}{\partial \theta} = \sum_\alpha\sum_{\beta\neq\alpha} \left[\frac{\partial w_{\alpha}}{\partial\theta} w_{\beta} + w_{\alpha}\frac{\partial w_{\beta}}{\partial\theta} \right] {\cal B}_{i}(x_{\alpha}, x_{\beta})\nonumber,
\eeq
and
\beq
\frac{\partial RR_i}{\partial \theta} &=& \left(\sum_\alpha \left[\frac{\partial w_\alpha}{\partial\theta} \left(\sum_\beta w_\beta - w_\alpha\right) \right. \right.  \\
&& \ \ \ \ \ \ \ \ \ \ \ \left. \left. + w_\alpha \left(\sum_\beta \frac{\partial w_\beta}{\partial\theta} - \frac{\partial w_\alpha}{\partial\theta}\right) \right]\right) \times \frac{V_i}{V}\nonumber
\eeq
We assume here that the parameters $\theta$ exclusively impact the value of the weights mapped onto each object. Should the object {\it positions} also depend on $\theta$, additional terms proportional to the spatial derivative of the bin selection function ${\cal B}_{i}(x_{\alpha}, x_{\beta})$ would arise. We leave the treatment of such contributions to the derivative to future work exploring models that parameterize changes to galaxy position. 

We note that the expressions derived in this section can be quite useful in cases where one has existing weighted pair-counting code to efficiently compute two-point functions. In this case, an autodiff library such as JAX can be used to calculate the per-object weights, $w_{\alpha},$ and the gradients  $\partial w_{\alpha}/\partial\theta;$ subsequently, the pair-counting code can proceed with its usual computation, using the equations derived in this section to transform the results into a computation of a two-point function and its exact derivatives. In carrying out the calculations in this section, we use used the {\tt corrfunc} code for this purpose. 

In the top panel of Figure~\ref{fig:wp_grads}, we show $\wprp$ for galaxies with $\Mstar>10^{11}\msun$ based on the BPL simulation, with the same fiducial scatter model and perturbations used in Fig.~\ref{fig:ds_grads}. In the bottom panel of Figure~\ref{fig:wp_grads}, we show gradients of $\wprp$ with respect to $\sigma_{\rm lo}$ and $\sigma_{\rm hi}.$ We see the same trends in Figure \ref{fig:wp_grads} as we saw in Fig. \ref{fig:ds_grads}: increasing scatter in the SMHM decreases clustering strength, with more pronounced effects from scatter at high mass.

\section{Discussion \& Future Work}
\label{sec:discussion}

We have presented a new approach to making simulation-based predictions of the galaxy--halo connection. In the conventional methodology, synthetic galaxy populations are generated from a stochastic Monte Carlo realization of some probability distribution defined by the underlying model. For example, in most abundance matching studies, stellar masses are drawn from a realization of a log-normal PDF; the synthetic galaxy at the center of each subhalo is then assigned to a unique bin of stellar mass (or otherwise discarded from the sample), and predictions for clustering and lensing are made from the resulting bins of point data \citep[as in, e.g.,][]{reddick_etal13,hearin_etal13}. By contrast, the approach taken here is not based on Monte Carlo realizations. Instead, {\em every} subhalo makes a PDF-weighted contribution to {\em each} stellar mass bin, and the statistical estimators used to calculate summary statistics are defined in terms of the weighted point data. 

Previous implementations of the galaxy--halo connection have also opted in favor of weighted summary statistics over hard-edged bins filled with noisy Monte Carlo realizations. For example, a variation on this technique was used in \citet{reid2014} in their treatment of fiber collisions; the {\tt UniverseMachine} source code used PDF-weighting methods in the computation of its DR1 summary statistics \citep{behroozi_etal19}; these same techniques form the basis of the {\tt TabCorr} source code\footnote{https://github.com/johannesulf/TabCorr}, as well as a variety of previous works on the galaxy--halo connection that utilize pre-computation methods \citep[e.g.,][]{neistein_etal12,zheng_guo_2016}. 

Although implementations of the galaxy--halo connection utilizing pre-computation methods also enjoy the same reduction of stochasticity as the approach presented here, our framework has two distinct advantages that make it more extensible to the problem sizes of cosmological modeling in the 2020s. First, conventional pre-computation methods are implemented by tabulating a set of results over a grid of (sub)halo properties, and then linearly interpolating from the resulting lookup table. While this methodology has proven to be quite effective in deriving constraints on cosmological parameters \citep[e.g.,][]{lange_etal21}, extending these techniques to models that depend upon more than one or two (sub)halo properties would be quite challenging, since implementations based on lookup-table interpolation have memory demands that rise sharply with the dimension of the table. Our use of AI resolves this issue of excessive memory demands, in essence by storing the lookup-table information in the weights and biases of a neural network. Second, our predictions for large scale structure are differentiable. Although we achieve this property in part due to our use of PDF-weighted summary statistics, as discussed above this is not the novel feature of our methodology. As described in \S\ref{sec:diffobs}, our predictions for the stellar mass function, galaxy--galaxy lensing, and galaxy clustering are differentiable as a result of the analytical propagation of the gradients of each synthetic galaxy's weight to the point estimators of the summary statistics. The availability of gradient information is the central feature that enables modern Bayesian inference techniques such as Hamiltonian Monte Carlo to derive converged posteriors for models of hundreds of parameters \citep{hoffman_gelman_2014_nuts}, and the extensibility of our framework to physical models of higher dimension is one of the principal motivations underlying this paper.

Our paper focuses on the most widely-used 1- and 2-point functions in cosmology, but our framework could naturally extend to higher-order summary statistics. In some cases, virtually no adaptation of the computations detailed in \S\ref{sec:diffobs} would be required. For example, both the under-density probability function (UPF) and counts-in-cells (CIC) summary statistics have been shown to be sensitive probes of galaxy--halo information that is not contained in galaxy clustering or lensing \citep{tinker_etal08,wang_etal19}. Both of these summary statistics are defined in terms of number counts of galaxies residing within some enclosing volume. If the UPF and CIC estimators are instead defined in terms of the total sum of galaxy weights residing in the enclosing volume, the predictions for these statistics become differentiable using the same techniques in \S\ref{sec:diffobs}. Recent work has shown that the information content of $n-$point functions of arbitrarily high order is formally contained in summary statistics based on $k$ nearest-neighbor ($k$NN) computations \citep{banerjee_abel_2020,banerjee_abel_2021}; this approach to extracting information from the density field is an attractive alternative to using higher-order $n-$point functions directly, since the computation of $k$NN-based summary statistics scales like $\mathcal{O}(N\log N);$ moreover, it has also recently been shown that $k$NN-based measurements are sensitive probes of the galaxy--halo connection \citep{behroozi_etal21}. In the standard computation of these summary statistics, the estimator is defined in terms of the spatial distance to the $N^{\rm th}-$nearest member of a galaxy sample defined by some hard-edged bin or threshold; in the corresponding differentiable version, one would instead use the smallest distance for which the sum of galaxy weights exceeds some chosen value. We relegate the extension of our framework to these and other higher-order summary statistics to future work.

Our work is closely related to \citet{horowitz_etal21}, who have implemented DiffHOD, a differentiable form of the Halo Occupation Distribution (HOD). Whereas the HOD populates {\it host} dark matter halos with variable numbers of satellites, SHAM populates each simulated {\it subhalo} with a single synthetic galaxy. Nonetheless, the theoretical roots of these two empirical models are tightly connected \citep[][]{kravtsov_etal04}, and DiffHOD and \shamnet are part of a growing trend of differentiable formulations of galaxy--halo connection models. In these two examples, the differentiability of the model predictions is achieved through a different methodology. In \shamnet, the derivatives of the galaxy--halo connection model are propagated through analytical PDFs of the galaxy residing in each halo, so that \shamnet predictions have no stochasticity; by contrast, DiffHOD uses a differentiable form of stochastic sampling based on a Gumbel-Softmax distribution, coupled with an annealing technique \citep{jang_gu_poole_gumbel_softmax}. 

The GalaxyNet model \citep{moster_2020_galaxynet} is also a differentiable form of the galaxy--halo connection, and so the techniques presented here could naturally be used to transform GalaxyNet predictions for clustering and lensing into differentiable computations. Whereas DiffHOD and \shamnet use AI-based methods in a manner that mimics the scaling relations of simple empirical models, GalaxyNet builds a highly nonlinear galaxy--halo connection with its neural network, and so is a far more expansive application of AI. Our two-phase training of \shamnet is an adaption of the same technique used to train GalaxyNet. As detailed in Appendix~\ref{sec:shamnet_train}, we trained \shamnet by first carrying out an initialization phase in which the weights and biases of the network were tuned to reproduce an existing model; training then proceeded with a subsequent phase in which the cost function was defined directly in terms of target summary statistics of the galaxy distribution. The initialization phase of \shamnet was based on a flexible parametric form of the stellar-to-halo-mass relation (see Appendix~\ref{sec:threeroll}), whereas GalaxyNet was initialized based on the EMERGE model of star formation history \citep{moster_etal2018_emerge}.

Our primary interest in the differentiable techniques introduced here is their application to more physically complex and higher-dimensional models than abundance matching. In particular, in closely related work \citep{hearin_etal21}, we have introduced {\tt diffmah}, a differentiable model for the assembly of dark matter halos; the {\tt diffmah} model is the basis of a new approach to the galaxy--halo connection, {\tt diffstar}, in which parameterized star formation histories (SFH) are statistically mapped onto the merger trees of dark matter halos \citep{alarcon_etal21}. Formulating the galaxy--halo connection in terms of parameterized SFHs allows us to additionally forward model galaxy SEDs in a differentiable fashion through the use of \href{https://github.com/ArgonneCPAC/dsps}{DSPS}, a JAX-based implementation of stellar population synthesis \citep{hearin_etal21_dsps}. Our focus on the benefits of gradient information in likelihood analyses anticipates the expansion of the dimension of the parameter space required by the complexity of these models.

In the present work, we have used SHAM primarily as a toy model to demonstrate how gradients of parameters of a galaxy--halo connection model propagate through to derivatives of predictions for the summary statistics of large scale structure. However, analyzing the galaxy distribution with SHAM is an active area of research unto itself, and \shamnet offers some improvements upon conventional deconvolution-based abundance matching that may be useful for such purposes. For example, the RL deconvolution algorithm outlined in \S\ref{sec:sham} has notoriously finicky convergence properties at the high-mass end, whereas \shamnet provides a robust solution to Eq.~\ref{eq:shamscatter} even at very high mass (see Figure~\ref{fig:shamnet_accuracy}). More importantly, \shamnet naturally incorporates mass-dependent scatter in the stellar-to-halo mass relation, whereas all previous deconvolution-based SHAM constraints have been derived under the assumption of constant scatter, at least in part due to the technical challenge of incorporating this feature into the RL-deconvolution algorithm.

With only modest extensions to \shamnet, differentiable versions of contemporary analyses could be applied to observational data. The present work is based on $\mpeak-$based abundance matching, and so in order to support abundance matching based on alternative mass proxies such as $V_{\rm peak}$ \citep{reddick_etal13}, $V_{\rm relax}$ \citep{chaves_montero_etal16}, or more generalized proxies \citep{lehmann_etal17}, it would be necessary to retrain \shamnet based on an alternative parameterization of the subhalo abundance function. Although in \S\ref{sec:diffobs} we focused on the gradients of the mass-dependent scatter, we trained \shamnet to additionally capture the dependence of $\smhm$ upon parameters specifying the SMF, and so our approach could also be used in abundance matching formulations such as \citet{saito_etal16} that incorporate uncertainty in the SMF. In order to incorporate cosmology-dependence into abundance matching, as in \citet{contreras_etal21}, an additional model providing a mapping from cosmological parameters to the parameterized subhalo abundance function would need to be developed. If an analysis required additional marginalization over uncertainty in orphan abundance, as in \citet{derose_etal21}, it would be necessary to develop an additional ingredient for how the parameters of the orphan prescription change the parameters of the subhalo abundance function.

With comparably modest adaptations, our framework could also be used to conduct differentiable analyses of conditional abundance matching \citep[CAM,][]{hearin_etal14}. While SHAM supplies a mapping from e.g., halo mass to stellar mass, the CAM framework supplies a mapping from some secondary subhalo property \citep[such as a halo formation time proxy, as in][]{hearin_watson_2013,masaki_etal13}, to some secondary galaxy property \citep[such as specific star formation rate, as in][]{watson_etal15}. In order to make CAM differentiable, rather than parameterizing the SMF and SHMF, one would instead need to calibrate parameterized models for the conditional abundance of the secondary halo and galaxy properties, and then use, e.g., the \href{https://halotools.readthedocs.io/en/latest/api/halotools.empirical_models.conditional_abunmatch.html}{conditional\_abunmatch} function in {\tt halotools} to generate training data for CAMNet. Although the version of \shamnet that we trained in the present work does not directly support these features, we have made our source code publicly available to facilitate these and other adaptations in future work.

\section{Summary}
\label{sec:conclusion}

\begin{enumerate}
    \item We have devised a new, differentiable framework for making simulation-based predictions for large scale structure. Our approach is not based on stochastic Monte Carlo realizations, but is instead formulated in terms of parameterized PDFs that are mapped onto each simulated halo. We then use automatic differentiation to propagate gradients of the galaxy--halo connection through the point estimators used to measure summary statistics of the galaxy distribution.
    \item We have used a neural network, \shamnet, to approximate the stellar-to-halo mass relationship (SMHM) of abundance matching. Our JAX-based implementation of \shamnet is available for installation with pip. In addition to being differentiable, \shamnet captures variable levels of $\mhalo$-dependent scatter in the SMHM, which is challenging to incorporate into conventional deconvolution-based implementations of abundance matching.
   \item Our source code is publicly available at \url{https://github.com/ArgonneCPAC/shamnet}, and provides a set of recipes that can be adapted to formulate other differentiable, AI-accelerated models of the galaxy--halo connection.
    \end{enumerate}

\section{Acknowledgements}
We thank Peter Behroozi for useful discussions. APH thanks José Feliciano again for Feliz Navidad.

We thank the developers of {\tt NumPy} \citep{numpy_ndarray}, {\tt SciPy} \citep{scipy}, Jupyter \citep{jupyter}, IPython \citep{ipython}, scikit-learn \citep{scikit_learn}, JAX \citep{jax2018github}, numba \citep{lam2015numba}, conda-forge \citep{conda_forge_community_2015_4774216}, and Matplotlib \citep{matplotlib} for their extremely useful free software. While writing this paper we made extensive use of the Astrophysics Data Service (ADS) and {\tt arXiv} preprint repository. The Bolshoi simulations used in this work have been performed within the Bolshoi project of the University of California High-Performance AstroComputing Center (UC-HiPACC) and were run at the NASA Ames Research Center.

Work done at Argonne National Laboratory was supported under the DOE contract DE-AC02-06CH11357. JD is supported by the Chamberlain Fellowship at Lawrence Berkeley National Laboratory. We gratefully acknowledge use of the Bebop cluster in the Laboratory Computing Resource Center at Argonne National Laboratory.



\bibliographystyle{aasjournal}
\bibliography{bibliography}

\begin{thebibliography}{}
\expandafter\ifx\csname natexlab\endcsname\relax\def\natexlab#1{#1}\fi
\providecommand{\url}[1]{\href{#1}{#1}}
\providecommand{\dodoi}[1]{doi:~\href{http://doi.org/#1}{\nolinkurl{#1}}}
\providecommand{\doeprint}[1]{\href{http://ascl.net/#1}{\nolinkurl{http://ascl.net/#1}}}
\providecommand{\doarXiv}[1]{\href{https://arxiv.org/abs/#1}{\nolinkurl{https://arxiv.org/abs/#1}}}

\bibitem[{{Alarcon} {et~al.}(2021){Alarcon}, {Hearin}, {Becker}, \&
  {Chaves-Montero}}]{alarcon_etal21}
{Alarcon}, A., {Hearin}, A., {Becker}, M., \& {Chaves-Montero}, J. 2021, in
  prep

\bibitem[{{Banerjee} \& {Abel}(2021{\natexlab{a}})}]{banerjee_abel_2020}
{Banerjee}, A., \& {Abel}, T. 2021{\natexlab{a}}, \mnras, 500, 5479,
  \dodoi{10.1093/mnras/staa3604}

\bibitem[{{Banerjee} \& {Abel}(2021{\natexlab{b}})}]{banerjee_abel_2021}
---. 2021{\natexlab{b}}, \mnras, 504, 2911, \dodoi{10.1093/mnras/stab961}

\bibitem[{{Behroozi} {et~al.}(2021){Behroozi}, {Hearin}, \&
  {Moster}}]{behroozi_etal21}
{Behroozi}, P., {Hearin}, A., \& {Moster}, B.~P. 2021, arXiv:2101.05280,
  arXiv:2101.05280.
\newblock \doarXiv{2101.05280}

\bibitem[{{Behroozi} {et~al.}(2019){Behroozi}, {Wechsler}, {Hearin}, \&
  {Conroy}}]{behroozi_etal19}
{Behroozi}, P., {Wechsler}, R.~H., {Hearin}, A.~P., \& {Conroy}, C. 2019,
  \mnras, 488, 3143, \dodoi{10.1093/mnras/stz1182}

\bibitem[{{Behroozi} {et~al.}(2010){Behroozi}, {Conroy}, \&
  {Wechsler}}]{behroozi_etal10}
{Behroozi}, P.~S., {Conroy}, C., \& {Wechsler}, R.~H. 2010, \apj, 717, 379,
  \dodoi{10.1088/0004-637X/717/1/379}

\bibitem[{{Behroozi} {et~al.}(2013{\natexlab{a}}){Behroozi}, {Wechsler}, \&
  {Conroy}}]{behroozi_etal13}
{Behroozi}, P.~S., {Wechsler}, R.~H., \& {Conroy}, C. 2013{\natexlab{a}}, ApJ,
  770, 57, \dodoi{10.1088/0004-637X/770/1/57}

\bibitem[{{Behroozi} {et~al.}(2013{\natexlab{b}}){Behroozi}, {Wechsler}, \&
  {Wu}}]{behroozi_etal13_rockstar}
{Behroozi}, P.~S., {Wechsler}, R.~H., \& {Wu}, H.-Y. 2013{\natexlab{b}}, \apj,
  762, 109, \dodoi{10.1088/0004-637X/762/2/109}

\bibitem[{{Behroozi} {et~al.}(2013{\natexlab{c}}){Behroozi}, {Wechsler}, {Wu},
  {Busha}, {Klypin}, \& {Primack}}]{behroozi_etal13_consistent_trees}
{Behroozi}, P.~S., {Wechsler}, R.~H., {Wu}, H.-Y., {et~al.} 2013{\natexlab{c}},
  \apj, 763, 18, \dodoi{10.1088/0004-637X/763/1/18}

\bibitem[{{Berlind} \& {Weinberg}(2002)}]{berlind_weinberg02}
{Berlind}, A.~A., \& {Weinberg}, D.~H. 2002, \apj, 575, 587,
  \dodoi{10.1086/341469}

\bibitem[{{Bernardeau} {et~al.}(2002){Bernardeau}, {Colombi}, {Gazta{\~n}aga},
  \& {Scoccimarro}}]{bernardeau_etal02}
{Bernardeau}, F., {Colombi}, S., {Gazta{\~n}aga}, E., \& {Scoccimarro}, R.
  2002, \physrep, 367, 1, \dodoi{10.1016/S0370-1573(02)00135-7}

\bibitem[{{Bocquet} {et~al.}(2020){Bocquet}, {Heitmann}, {Habib}, {Lawrence},
  {Uram}, {Frontiere}, {Pope}, \& {Finkel}}]{bocquet_etal20}
{Bocquet}, S., {Heitmann}, K., {Habib}, S., {et~al.} 2020, \apj, 901, 5,
  \dodoi{10.3847/1538-4357/abac5c}

\bibitem[{Bradbury {et~al.}(2018)Bradbury, Frostig, Hawkins, Johnson, Leary,
  Maclaurin, Necula, Paszke, Vander{P}las, Wanderman-{M}ilne, \&
  Zhang}]{jax2018github}
Bradbury, J., Frostig, R., Hawkins, P., {et~al.} 2018, {JAX}: composable
  transformations of {P}ython+{N}um{P}y programs, 0.2.10.
\newblock \url{http://github.com/google/jax}

\bibitem[{{Cacciato} {et~al.}(2013){Cacciato}, {van den Bosch}, {More}, {Mo},
  \& {Yang}}]{cacciato_etal13}
{Cacciato}, M., {van den Bosch}, F.~C., {More}, S., {Mo}, H., \& {Yang}, X.
  2013, \mnras, 430, 767, \dodoi{10.1093/mnras/sts525}

\bibitem[{{Campbell} {et~al.}(2018){Campbell}, {van den Bosch}, {Padmanabhan},
  {Mao}, {Zentner}, {Lange}, {Jiang}, \& {Villarreal}}]{campbell_etal18}
{Campbell}, D., {van den Bosch}, F.~C., {Padmanabhan}, N., {et~al.} 2018,
  \mnras, 477, 359, \dodoi{10.1093/mnras/sty495}

\bibitem[{{Carrasco} {et~al.}(2012){Carrasco}, {Hertzberg}, \&
  {Senatore}}]{carrasco_etal12}
{Carrasco}, J. J.~M., {Hertzberg}, M.~P., \& {Senatore}, L. 2012, Journal of
  High Energy Physics, 2012, 82, \dodoi{10.1007/JHEP09(2012)082}

\bibitem[{{Chapman} {et~al.}(2021){Chapman}, {Mohammad}, {Zhai}, {Percival},
  {Tinker}, {Bautista}, {Brownstein}, {Burtin}, {Dawson}, {Gil-Mar{\'\i}n}, {de
  la Macorra}, {Ross}, {Rossi}, {Schneider}, \& {Zhao}}]{chapman_etal21}
{Chapman}, M.~J., {Mohammad}, F.~G., {Zhai}, Z., {et~al.} 2021,
  arXiv:2106.14961, arXiv:2106.14961.
\newblock \doarXiv{2106.14961}

\bibitem[{{Chaves-Montero} {et~al.}(2016){Chaves-Montero}, {Angulo}, {Schaye},
  {Schaller}, {Crain}, {Furlong}, \& {Theuns}}]{chaves_montero_etal16}
{Chaves-Montero}, J., {Angulo}, R.~E., {Schaye}, J., {et~al.} 2016, \mnras,
  460, 3100, \dodoi{10.1093/mnras/stw1225}

\bibitem[{conda-forge community(2015)}]{conda_forge_community_2015_4774216}
conda-forge community. 2015, {The conda-forge Project: Community-based Software
  Distribution Built on the conda Package Format and Ecosystem},  Zenodo,
  \dodoi{10.5281/zenodo.4774216}

\bibitem[{{Conroy} {et~al.}(2006){Conroy}, {Wechsler}, \&
  {Kravtsov}}]{conroy_etal06}
{Conroy}, C., {Wechsler}, R.~H., \& {Kravtsov}, A.~V. 2006, \apj, 647, 201,
  \dodoi{10.1086/503602}

\bibitem[{{Contreras} {et~al.}(2021){Contreras}, {Chaves-Montero}, {Zennaro},
  \& {Angulo}}]{contreras_etal21}
{Contreras}, S., {Chaves-Montero}, J., {Zennaro}, M., \& {Angulo}, R.~E. 2021,
  arXiv e-prints, arXiv:2105.05854.
\newblock \doarXiv{2105.05854}

\bibitem[{{DeRose} {et~al.}(2021{\natexlab{a}}){DeRose}, {Becker}, \&
  {Wechsler}}]{derose_etal21}
{DeRose}, J., {Becker}, M.~R., \& {Wechsler}, R.~H. 2021{\natexlab{a}}, arXiv
  e-prints, arXiv:2105.12104.
\newblock \doarXiv{2105.12104}

\bibitem[{{DeRose} {et~al.}(2021{\natexlab{b}}){DeRose}, {Chen}, {White}, \&
  {Kokron}}]{derose_etal21b}
{DeRose}, J., {Chen}, S.-F., {White}, M., \& {Kokron}, N. 2021{\natexlab{b}},
  arXiv:2112.05889, arXiv:2112.05889.
\newblock \doarXiv{2112.05889}

\bibitem[{{DES Collaboration} {et~al.}(2021){DES Collaboration}, {Abbott},
  {Aguena}, {Alarcon}, {Allam}, {Alves}, {Amon}, {Andrade-Oliveira}, {Annis},
  {Avila}, {Bacon}, {Baxter}, {Bechtol}, {Becker}, {Bernstein}, {Bhargava},
  {Birrer}, {Blazek}, {Brandao-Souza}, {Bridle}, {Brooks}, {Buckley-Geer},
  {Burke}, {Camacho}, {Campos}, {Carnero Rosell}, {Carrasco Kind}, {Carretero},
  {Castander}, {Cawthon}, {Chang}, {Chen}, {Chen}, {Choi}, {Conselice},
  {Cordero}, {Costanzi}, {Crocce}, {da Costa}, {da Silva Pereira}, {Davis},
  {Davis}, {De Vicente}, {DeRose}, {Desai}, {Di Valentino}, {Diehl},
  {Dietrich}, {Dodelson}, {Doel}, {Doux}, {Drlica-Wagner}, {Eckert}, {Eifler},
  {Elsner}, {Elvin-Poole}, {Everett}, {Evrard}, {Fang}, {Farahi}, {Fernandez},
  {Ferrero}, {Fert{\'e}}, {Fosalba}, {Friedrich}, {Frieman},
  {Garc{\'\i}a-Bellido}, {Gatti}, {Gaztanaga}, {Gerdes}, {Giannantonio},
  {Giannini}, {Gruen}, {Gruendl}, {Gschwend}, {Gutierrez}, {Harrison},
  {Hartley}, {Herner}, {Hinton}, {Hollowood}, {Honscheid}, {Hoyle}, {Huff},
  {Huterer}, {Jain}, {James}, {Jarvis}, {Jeffrey}, {Jeltema}, {Kovacs},
  {Krause}, {Kron}, {Kuehn}, {Kuropatkin}, {Lahav}, {Leget}, {Lemos}, {Liddle},
  {Lidman}, {Lima}, {Lin}, {MacCrann}, {Maia}, {Marshall}, {Martini},
  {McCullough}, {Melchior}, {Mena-Fern{\'a}ndez}, {Menanteau}, {Miquel},
  {Mohr}, {Morgan}, {Muir}, {Myles}, {Nadathur}, {Navarro-Alsina}, {Nichol},
  {Ogando}, {Omori}, {Palmese}, {Pandey}, {Park}, {Paz-Chinch{\'o}n},
  {Petravick}, {Pieres}, {Plazas Malag{\'o}n}, {Porredon}, {Prat}, {Raveri},
  {Rodriguez-Monroy}, {Rollins}, {Romer}, {Roodman}, {Rosenfeld}, {Ross},
  {Rykoff}, {Samuroff}, {S{\'a}nchez}, {Sanchez}, {Sanchez}, {Sanchez Cid},
  {Scarpine}, {Schubnell}, {Scolnic}, {Secco}, {Serrano}, {Sevilla-Noarbe},
  {Sheldon}, {Shin}, {Smith}, {Soares-Santos}, {Suchyta}, {Swanson}, {Tabbutt},
  {Tarle}, {Thomas}, {To}, {Troja}, {Troxel}, {Tucker}, {Tutusaus}, {Varga},
  {Walker}, {Weaverdyck}, {Weller}, {Yanny}, {Yin}, {Zhang}, \&
  {Zuntz}}]{des_y3_3_by_2}
{DES Collaboration}, {Abbott}, T.~M.~C., {Aguena}, M., {et~al.} 2021, arXiv
  e-prints, arXiv:2105.13549.
\newblock \doarXiv{2105.13549}

\bibitem[{{Desjacques} {et~al.}(2018){Desjacques}, {Jeong}, \&
  {Schmidt}}]{desjacques_etal18}
{Desjacques}, V., {Jeong}, D., \& {Schmidt}, F. 2018, \physrep, 733, 1,
  \dodoi{10.1016/j.physrep.2017.12.002}

\bibitem[{{Eddington}(1913)}]{Eddington13}
{Eddington}, A.~S. 1913, MNRAS, 73, 359, \dodoi{10.1093/mnras/73.5.359}

\bibitem[{{Euclid Collaboration} {et~al.}(2020){Euclid Collaboration},
  {Knabenhans}, {Stadel}, {Potter}, {Dakin}, {Hannestad}, {Tram}, {Marelli},
  {Schneider}, {Teyssier}, {Andreon}, {Auricchio}, {Baccigalupi},
  {Balaguera-Antol{\'\i}nez}, {Baldi}, {Bardelli}, {Battaglia}, {Bender},
  {Biviano}, {Bodendorf}, {Bozzo}, {Branchini}, {Brescia}, {Burigana},
  {Cabanac}, {Camera}, {Capobianco}, {Cappi}, {Carbone}, {Carretero},
  {Carvalho}, {Casas}, {Casas}, {Castellano}, {Castignani}, {Cavuoti},
  {Cledassou}, {Colodro-Conde}, {Congedo}, {Conselice}, {Conversi}, {Copin},
  {Corcione}, {Coupon}, {Courtois}, {Da Silva}, {de la Torre}, {Di Ferdinando},
  {Duncan}, {Dupac}, {Fabbian}, {Farrens}, {Ferreira}, {Finelli}, {Frailis},
  {Franceschi}, {Galeotta}, {Garilli}, {Giocoli}, {Gozaliasl},
  {Graci{\'a}-Carpio}, {Grupp}, {Guzzo}, {Holmes}, {Hormuth}, {Israel},
  {Jahnke}, {Keihanen}, {Kermiche}, {Kirkpatrick}, {Kubik}, {Kunz},
  {Kurki-Suonio}, {Ligori}, {Lilje}, {Lloro}, {Maino}, {Marggraf}, {Markovic},
  {Martinet}, {Marulli}, {Massey}, {Mauri}, {Maurogordato}, {Medinaceli},
  {Meneghetti}, {Metcalf}, {Meylan}, {Moresco}, {Morin}, {Moscardini},
  {Munari}, {Neissner}, {Niemi}, {Padilla}, {Paltani}, {Pasian}, {Patrizii},
  {Pettorino}, {Pires}, {Polenta}, {Poncet}, {Raison}, {Renzi}, {Rhodes},
  {Riccio}, {Romelli}, {Roncarelli}, {Saglia}, {S{\'a}nchez}, {Sapone},
  {Schneider}, {Scottez}, {Secroun}, {Serrano}, {Sirignano}, {Sirri}, {Stanco},
  {Sureau}, {Tallada Cresp{\'\i}}, {Taylor}, {Tenti}, {Tereno}, {Toledo-Moreo},
  {Torradeflot}, {Valenziano}, {Valiviita}, {Vassallo}, {Viel}, {Wang},
  {Welikala}, {Whittaker}, {Zacchei}, \& {Zucca}}]{euclid_emulator_pk_2020}
{Euclid Collaboration}, {Knabenhans}, M., {Stadel}, J., {et~al.} 2020,
  arXiv:2010.11288, arXiv:2010.11288.
\newblock \doarXiv{2010.11288}

\bibitem[{{Foreman-Mackey} {et~al.}(2019){Foreman-Mackey}, {Farr}, {Sinha},
  {Archibald}, {Hogg}, {Sanders}, {Zuntz}, {Williams}, {Nelson}, {de
  Val-Borro}, {Erhardt}, {Pashchenko}, \& {Pla}}]{emcee_2019}
{Foreman-Mackey}, D., {Farr}, W., {Sinha}, M., {et~al.} 2019, The Journal of
  Open Source Software, 4, 1864, \dodoi{10.21105/joss.01864}

\bibitem[{{Garc{\'\i}a} {et~al.}(2021){Garc{\'\i}a}, {Rozo}, {Becker}, \&
  {More}}]{garcia_etal21_halo_exclusion}
{Garc{\'\i}a}, R., {Rozo}, E., {Becker}, M.~R., \& {More}, S. 2021, \mnras,
  \dodoi{10.1093/mnras/stab1317}

\bibitem[{{Hearin} {et~al.}(2021{\natexlab{a}}){Hearin}, {Chaves-Montero},
  {Alarcon}, {Becker}, \& {Benson}}]{hearin_etal21_dsps}
{Hearin}, A.~P., {Chaves-Montero}, J., {Alarcon}, A., {Becker}, M.~R., \&
  {Benson}, A. 2021{\natexlab{a}}, arXiv:2112.06830, arXiv:2112.06830.
\newblock \doarXiv{2112.06830}

\bibitem[{{Hearin} {et~al.}(2021{\natexlab{b}}){Hearin}, {Chaves-Montero},
  {Becker}, \& {Alarcon}}]{hearin_etal21}
{Hearin}, A.~P., {Chaves-Montero}, J., {Becker}, M.~R., \& {Alarcon}, A.
  2021{\natexlab{b}}, arXiv e-prints, arXiv:2105.05859.
\newblock \doarXiv{2105.05859}

\bibitem[{{Hearin} \& {Watson}(2013)}]{hearin_watson_2013}
{Hearin}, A.~P., \& {Watson}, D.~F. 2013, \mnras, 435, 1313,
  \dodoi{10.1093/mnras/stt1374}

\bibitem[{{Hearin} {et~al.}(2014){Hearin}, {Watson}, {Becker}, {Reyes},
  {Berlind}, \& {Zentner}}]{hearin_etal14}
{Hearin}, A.~P., {Watson}, D.~F., {Becker}, M.~R., {et~al.} 2014, \mnras, 444,
  729, \dodoi{10.1093/mnras/stu1443}

\bibitem[{{Hearin} {et~al.}(2013){Hearin}, {Zentner}, {Berlind}, \&
  {Newman}}]{hearin_etal13}
{Hearin}, A.~P., {Zentner}, A.~R., {Berlind}, A.~A., \& {Newman}, J.~A. 2013,
  \mnras, 433, 659, \dodoi{10.1093/mnras/stt755}

\bibitem[{{Hearin} {et~al.}(2017){Hearin}, {Campbell}, {Tollerud}, {Behroozi},
  {Diemer}, {Goldbaum}, {Jennings}, {Leauthaud}, {Mao}, {More}, {Parejko},
  {Sinha}, {Sip{\"o}cz}, \& {Zentner}}]{hearin_etal17_halotools}
{Hearin}, A.~P., {Campbell}, D., {Tollerud}, E., {et~al.} 2017, \aj, 154, 190,
  \dodoi{10.3847/1538-3881/aa859f}

\bibitem[{{Heitmann} {et~al.}(2006){Heitmann}, {Higdon}, {Nakhleh}, \&
  {Habib}}]{heitmann_etal06}
{Heitmann}, K., {Higdon}, D., {Nakhleh}, C., \& {Habib}, S. 2006, \apjl, 646,
  L1, \dodoi{10.1086/506448}

\bibitem[{{Hoffman} \& {Gelman}(2014)}]{hoffman_gelman_2014_nuts}
{Hoffman}, M., \& {Gelman}, A. 2014, Journal of Machine Learning Research, 15,
  1593.
\newblock \url{http://jmlr.org/papers/v15/hoffman14a.html}

\bibitem[{{Horowitz} {et~al.}(2021){Horowitz}, {Hahn}, {Lanusse}, {Modi}, \&
  {Ferraro}}]{horowitz_etal21}
{Horowitz}, B., {Hahn}, C., {Lanusse}, F., {Modi}, C., \& {Ferraro}, S. 2021,
  in prep

\bibitem[{Hunter(2007)}]{matplotlib}
Hunter, J.~D. 2007, Computing In Science \& Engineering, 9, 90,
  \dodoi{10.1109/MCSE.2007.55}

\bibitem[{{Iman} {et~al.}(1981){Iman}, {Helton}, \&
  {Campbell}}]{iman_helton_campbell_1981_latin_hypercubes}
{Iman}, R., {Helton}, J., \& {Campbell}, J. 1981, Journal of Quality
  Technology, 13, 174, \dodoi{10.1080/00224065.1981.11978748}

\bibitem[{{Jang} {et~al.}(2016){Jang}, {Gu}, \&
  {Poole}}]{jang_gu_poole_gumbel_softmax}
{Jang}, E., {Gu}, S., \& {Poole}, B. 2016, arXiv:1611.01144, arXiv:1611.01144.
\newblock \doarXiv{1611.01144}

\bibitem[{{Jenkins} {et~al.}(2001){Jenkins}, {Frenk}, {White}, {Colberg},
  {Cole}, {Evrard}, {Couchman}, \& {Yoshida}}]{jenkins_etal01}
{Jenkins}, A., {Frenk}, C.~S., {White}, S.~D.~M., {et~al.} 2001, \mnras, 321,
  372, \dodoi{10.1046/j.1365-8711.2001.04029.x}

\bibitem[{Jones {et~al.}(2001-2016)Jones, Oliphant, Peterson, {et~al.}}]{scipy}
Jones, E., Oliphant, T., Peterson, P., {et~al.} 2001-2016, http://www.scipy.org

\bibitem[{{Joudaki} {et~al.}(2018){Joudaki}, {Blake}, {Johnson}, {Amon},
  {Asgari}, {Choi}, {Erben}, {Glazebrook}, {Harnois-D{\'e}raps}, {Heymans},
  {Hildebrandt}, {Hoekstra}, {Klaes}, {Kuijken}, {Lidman}, {Mead}, {Miller},
  {Parkinson}, {Poole}, {Schneider}, {Viola}, \& {Wolf}}]{joudaki_etal18}
{Joudaki}, S., {Blake}, C., {Johnson}, A., {et~al.} 2018, \mnras, 474, 4894,
  \dodoi{10.1093/mnras/stx2820}

\bibitem[{{Kingma} \& {Ba}(2014)}]{kingma_ba_adam_2015}
{Kingma}, D.~P., \& {Ba}, J. 2014, arXiv:1412.6980, arXiv:1412.6980.
\newblock \doarXiv{1412.6980}

\bibitem[{{Klambauer} {et~al.}(2017){Klambauer}, {Unterthiner}, {Mayr}, \&
  {Hochreiter}}]{klambauer_etal17}
{Klambauer}, G., {Unterthiner}, T., {Mayr}, A., \& {Hochreiter}, S. 2017,
  arXiv:1706.02515, arXiv:1706.02515.
\newblock \doarXiv{1706.02515}

\bibitem[{{Klypin} {et~al.}(2011){Klypin}, {Trujillo-Gomez}, \&
  {Primack}}]{klypin_etal11}
{Klypin}, A.~A., {Trujillo-Gomez}, S., \& {Primack}, J. 2011, \apj, 740, 102,
  \dodoi{10.1088/0004-637X/740/2/102}

\bibitem[{{Kokron} {et~al.}(2021){Kokron}, {DeRose}, {Chen}, {White}, \&
  {Wechsler}}]{kokron_etal21}
{Kokron}, N., {DeRose}, J., {Chen}, S.-F., {White}, M., \& {Wechsler}, R.~H.
  2021, \mnras, 505, 1422, \dodoi{10.1093/mnras/stab1358}

\bibitem[{{Krause} \& {Eifler}(2017)}]{krause_etal17}
{Krause}, E., \& {Eifler}, T. 2017, \mnras, 470, 2100,
  \dodoi{10.1093/mnras/stx1261}

\bibitem[{{Kravtsov} {et~al.}(2004){Kravtsov}, {Berlind}, {Wechsler}, {Klypin},
  {Gottl{\"o}ber}, {Allgood}, \& {Primack}}]{kravtsov_etal04}
{Kravtsov}, A.~V., {Berlind}, A.~A., {Wechsler}, R.~H., {et~al.} 2004, \apj,
  609, 35, \dodoi{10.1086/420959}

\bibitem[{{Kravtsov} {et~al.}(1997){Kravtsov}, {Klypin}, \&
  {Khokhlov}}]{kravtsov_etal97_art}
{Kravtsov}, A.~V., {Klypin}, A.~A., \& {Khokhlov}, A.~M. 1997, \apjs, 111, 73,
  \dodoi{10.1086/313015}

\bibitem[{{Kravtsov} {et~al.}(2018){Kravtsov}, {Vikhlinin}, \&
  {Meshcheryakov}}]{kravtsov_etal18}
{Kravtsov}, A.~V., {Vikhlinin}, A.~A., \& {Meshcheryakov}, A.~V. 2018,
  Astronomy Letters, 44, 8, \dodoi{10.1134/S1063773717120015}

\bibitem[{{Kwan} {et~al.}(2015){Kwan}, {Heitmann}, {Habib}, {Padmanabhan},
  {Lawrence}, {Finkel}, {Frontiere}, \& {Pope}}]{kwan_etal15}
{Kwan}, J., {Heitmann}, K., {Habib}, S., {et~al.} 2015, \apj, 810, 35,
  \dodoi{10.1088/0004-637X/810/1/35}

\bibitem[{Lam {et~al.}(2015)Lam, Pitrou, \& Seibert}]{lam2015numba}
Lam, S.~K., Pitrou, A., \& Seibert, S. 2015, in Proceedings of the Second
  Workshop on the LLVM Compiler Infrastructure in HPC, 1--6

\bibitem[{{Lange} {et~al.}(2021){Lange}, {Hearin}, {Leauthaud}, {van den
  Bosch}, {Guo}, \& {DeRose}}]{lange_etal21}
{Lange}, J.~U., {Hearin}, A.~P., {Leauthaud}, A., {et~al.} 2021,
  arXiv:2101.12261, arXiv:2101.12261.
\newblock \doarXiv{2101.12261}

\bibitem[{{Lange} {et~al.}(2019){Lange}, {Yang}, {Guo}, {Luo}, \& {van den
  Bosch}}]{lange_etal19}
{Lange}, J.~U., {Yang}, X., {Guo}, H., {Luo}, W., \& {van den Bosch}, F.~C.
  2019, MNRAS, 488, 5771, \dodoi{10.1093/mnras/stz2124}

\bibitem[{{Lawrence} {et~al.}(2017){Lawrence}, {Heitmann}, {Kwan}, {Upadhye},
  {Bingham}, {Habib}, {Higdon}, {Pope}, {Finkel}, \&
  {Frontiere}}]{lawrence_etal17}
{Lawrence}, E., {Heitmann}, K., {Kwan}, J., {et~al.} 2017, \apj, 847, 50,
  \dodoi{10.3847/1538-4357/aa86a9}

\bibitem[{{Lehmann} {et~al.}(2017){Lehmann}, {Mao}, {Becker}, {Skillman}, \&
  {Wechsler}}]{lehmann_etal17}
{Lehmann}, B.~V., {Mao}, Y.-Y., {Becker}, M.~R., {Skillman}, S.~W., \&
  {Wechsler}, R.~H. 2017, \apj, 834, 37, \dodoi{10.3847/1538-4357/834/1/37}

\bibitem[{{Li} \& {White}(2009)}]{li_white_2009}
{Li}, C., \& {White}, S. D.~M. 2009, \mnras, 398, 2177,
  \dodoi{10.1111/j.1365-2966.2009.15268.x}

\bibitem[{{Lucy}(1974)}]{lucy74}
{Lucy}, L.~B. 1974, AJ, 79, 745, \dodoi{10.1086/111605}

\bibitem[{{Lupton} {et~al.}(1999){Lupton}, {Gunn}, \&
  {Szalay}}]{lupton_gunn_szalay99}
{Lupton}, R.~H., {Gunn}, J.~E., \& {Szalay}, A.~S. 1999, \aj, 118, 1406,
  \dodoi{10.1086/301004}

\bibitem[{{Masaki} {et~al.}(2013){Masaki}, {Lin}, \& {Yoshida}}]{masaki_etal13}
{Masaki}, S., {Lin}, Y.-T., \& {Yoshida}, N. 2013, \mnras, 436, 2286,
  \dodoi{10.1093/mnras/stt1729}

\bibitem[{{McClintock} {et~al.}(2019{\natexlab{a}}){McClintock}, {Rozo},
  {Becker}, {DeRose}, {Mao}, {McLaughlin}, {Tinker}, {Wechsler}, \&
  {Zhai}}]{mcclintock_etal19}
{McClintock}, T., {Rozo}, E., {Becker}, M.~R., {et~al.} 2019{\natexlab{a}},
  \apj, 872, 53, \dodoi{10.3847/1538-4357/aaf568}

\bibitem[{{McClintock} {et~al.}(2019{\natexlab{b}}){McClintock}, {Rozo},
  {Banerjee}, {Becker}, {DeRose}, {McLaughlin}, {Tinker}, {Wechsler}, \&
  {Zhai}}]{mcclintock_etal20}
{McClintock}, T., {Rozo}, E., {Banerjee}, A., {et~al.} 2019{\natexlab{b}},
  arXiv:1907.13167, arXiv:1907.13167.
\newblock \doarXiv{1907.13167}

\bibitem[{{McKay} {et~al.}(1979){McKay}, {Beckman}, \&
  {Conover}}]{mckay_beckman_conover_1979_latin_hypercubes}
{McKay}, M., {Beckman}, R., \& {Conover}, W. 1979, Technometrics, 21, 239,
  \dodoi{10.1080/00401706.1979.10489755}

\bibitem[{{Miyatake} {et~al.}(2021){Miyatake}, {Sugiyama}, {Takada},
  {Nishimichi}, {Shirasaki}, {Kobayashi}, {Mandelbaum}, {More}, {Oguri},
  {Osato}, {Park}, {Takahashi}, {Coupon}, {Hikage}, {Hsieh}, {Leauthaud}, {Li},
  {Luo}, {Lupton}, {Miyazaki}, {Murayama}, {Nishizawa}, {Price}, {Simet},
  {Speagle}, {Strauss}, {Tanaka}, \& {Yoshida}}]{miyatake_etal21}
{Miyatake}, H., {Sugiyama}, S., {Takada}, M., {et~al.} 2021, arXiv:2111.02419,
  arXiv:2111.02419.
\newblock \doarXiv{2111.02419}

\bibitem[{{Moster} {et~al.}(2020){Moster}, {Naab}, {Lindstr{\"o}m}, \&
  {O'Leary}}]{moster_2020_galaxynet}
{Moster}, B.~P., {Naab}, T., {Lindstr{\"o}m}, M., \& {O'Leary}, J.~A. 2020,
  arXiv:2005.12276, arXiv:2005.12276.
\newblock \doarXiv{2005.12276}

\bibitem[{{Moster} {et~al.}(2013){Moster}, {Naab}, \& {White}}]{moster_etal13}
{Moster}, B.~P., {Naab}, T., \& {White}, S. D.~M. 2013, \mnras, 428, 3121,
  \dodoi{10.1093/mnras/sts261}

\bibitem[{{Moster} {et~al.}(2018){Moster}, {Naab}, \&
  {White}}]{moster_etal2018_emerge}
---. 2018, \mnras, 477, 1822, \dodoi{10.1093/mnras/sty655}

\bibitem[{{Moster} {et~al.}(2010){Moster}, {Somerville}, {Maulbetsch}, {van den
  Bosch}, {Macci{\`o}}, {Naab}, \& {Oser}}]{moster_etal10}
{Moster}, B.~P., {Somerville}, R.~S., {Maulbetsch}, C., {et~al.} 2010, \apj,
  710, 903, \dodoi{10.1088/0004-637X/710/2/903}

\bibitem[{{Neistein} \& {Khochfar}(2012)}]{neistein_etal12}
{Neistein}, E., \& {Khochfar}, S. 2012, arXiv:1209.0463, arXiv:1209.0463.
\newblock \doarXiv{1209.0463}

\bibitem[{{Nishimichi} {et~al.}(2019){Nishimichi}, {Takada}, {Takahashi},
  {Osato}, {Shirasaki}, {Oogi}, {Miyatake}, {Oguri}, {Murata}, {Kobayashi}, \&
  {Yoshida}}]{nishimichi_etal19}
{Nishimichi}, T., {Takada}, M., {Takahashi}, R., {et~al.} 2019, \apj, 884, 29,
  \dodoi{10.3847/1538-4357/ab3719}

\bibitem[{{Panter} {et~al.}(2007){Panter}, {Jimenez}, {Heavens}, \&
  {Charlot}}]{panter_etal07}
{Panter}, B., {Jimenez}, R., {Heavens}, A.~F., \& {Charlot}, S. 2007, \mnras,
  378, 1550, \dodoi{10.1111/j.1365-2966.2007.11909.x}

\bibitem[{Pedregosa {et~al.}(2011)Pedregosa, Varoquaux, Gramfort, Michel,
  Thirion, Grisel, Blondel, Prettenhofer, Weiss, Dubourg, Vanderplas, Passos,
  Cournapeau, Brucher, Perrot, \& Duchesnay}]{scikit_learn}
Pedregosa, F., Varoquaux, G., Gramfort, A., {et~al.} 2011, Journal of Machine
  Learning Research, 12, 2825

\bibitem[{P\'erez \& Granger(2007)}]{ipython}
P\'erez, F., \& Granger, B.~E. 2007, Computing in Science and Engineering, 9,
  21, \dodoi{10.1109/MCSE.2007.53}

\bibitem[{{Planck Collaboration} {et~al.}(2014){Planck Collaboration}, {Ade},
  {Aghanim}, {Armitage-Caplan}, {Arnaud}, {Ashdown}, {Atrio-Barandela},
  {Aumont}, {Baccigalupi}, {Banday}, \& et~al.}]{planck14b}
{Planck Collaboration}, {Ade}, P.~A.~R., {Aghanim}, N., {et~al.} 2014, \aap,
  571, A16, \dodoi{10.1051/0004-6361/201321591}

\bibitem[{{Press} \& {Schechter}(1974)}]{press_schechter_1974}
{Press}, W.~H., \& {Schechter}, P. 1974, \apj, 187, 425, \dodoi{10.1086/152650}

\bibitem[{{Ragan-Kelley} {et~al.}(2014){Ragan-Kelley}, {Perez}, {Granger},
  {Kluyver}, {Ivanov}, {Frederic}, \& {Bussonier}}]{jupyter}
{Ragan-Kelley}, M., {Perez}, F., {Granger}, B., {et~al.} 2014, in American
  Geophysical Union Fall Meeting Abstracts, Vol.~D7

\bibitem[{{Rasmussen} \& {Williams}(2006)}]{rasmussen_williams_GPbook_2006}
{Rasmussen}, C.~E., \& {Williams}, C. K.~I. 2006, {Gaussian Processes for
  Machine Learning}

\bibitem[{{Reddick} {et~al.}(2014){Reddick}, {Tinker}, {Wechsler}, \&
  {Lu}}]{reddick_etal14}
{Reddick}, R.~M., {Tinker}, J.~L., {Wechsler}, R.~H., \& {Lu}, Y. 2014, \apj,
  783, 118, \dodoi{10.1088/0004-637X/783/2/118}

\bibitem[{{Reddick} {et~al.}(2013){Reddick}, {Wechsler}, {Tinker}, \&
  {Behroozi}}]{reddick_etal13}
{Reddick}, R.~M., {Wechsler}, R.~H., {Tinker}, J.~L., \& {Behroozi}, P.~S.
  2013, \apj, 771, 30, \dodoi{10.1088/0004-637X/771/1/30}

\bibitem[{{Reid} {et~al.}(2014){Reid}, {Seo}, {Leauthaud}, {Tinker}, \&
  {White}}]{reid2014}
{Reid}, B.~A., {Seo}, H.-J., {Leauthaud}, A., {Tinker}, J.~L., \& {White}, M.
  2014, \mnras, 444, 476, \dodoi{10.1093/mnras/stu1391}

\bibitem[{Richardson(1972)}]{richardson72}
Richardson, W.~H. 1972, J. Opt. Soc. Am., 62, 55,
  \dodoi{10.1364/JOSA.62.000055}

\bibitem[{{Rodr{\'\i}guez-Puebla} {et~al.}(2015){Rodr{\'\i}guez-Puebla},
  {Avila-Reese}, {Yang}, {Foucaud}, {Drory}, \&
  {Jing}}]{rodriguez_puebla_etal15_smhm_colors}
{Rodr{\'\i}guez-Puebla}, A., {Avila-Reese}, V., {Yang}, X., {et~al.} 2015,
  \apj, 799, 130, \dodoi{10.1088/0004-637X/799/2/130}

\bibitem[{{Rodr{\'\i}guez-Puebla} {et~al.}(2016){Rodr{\'\i}guez-Puebla},
  {Behroozi}, {Primack}, {Klypin}, {Lee}, \&
  {Hellinger}}]{rodriguez_puebla_etal16}
{Rodr{\'\i}guez-Puebla}, A., {Behroozi}, P., {Primack}, J., {et~al.} 2016,
  \mnras, 462, 893, \dodoi{10.1093/mnras/stw1705}

\bibitem[{{Saito} {et~al.}(2016){Saito}, {Leauthaud}, {Hearin}, {Bundy},
  {Zentner}, {Behroozi}, {Reid}, {Sinha}, {Coupon}, {Tinker}, {White}, \&
  {Schneider}}]{saito_etal16}
{Saito}, S., {Leauthaud}, A., {Hearin}, A.~P., {et~al.} 2016, \mnras, 460,
  1457, \dodoi{10.1093/mnras/stw1080}

\bibitem[{{Sgr{\'o}} {et~al.}(2013){Sgr{\'o}}, {Paz}, \&
  {Merch{\'a}n}}]{sgro_etal13_satellite_anisotropy}
{Sgr{\'o}}, M.~A., {Paz}, D.~J., \& {Merch{\'a}n}, M. 2013, \mnras, 433, 787,
  \dodoi{10.1093/mnras/stt773}

\bibitem[{{Sinha} \& {Garrison}(2020)}]{corrfunc}
{Sinha}, M., \& {Garrison}, L.~H. 2020, \mnras, 491, 3022,
  \dodoi{10.1093/mnras/stz3157}

\bibitem[{{Tasitsiomi} {et~al.}(2004){Tasitsiomi}, {Kravtsov}, {Wechsler}, \&
  {Primack}}]{tasitsiomi_etal04}
{Tasitsiomi}, A., {Kravtsov}, A.~V., {Wechsler}, R.~H., \& {Primack}, J.~R.
  2004, \apj, 614, 533, \dodoi{10.1086/423784}

\bibitem[{{Tinker} {et~al.}(2008){Tinker}, {Conroy}, {Norberg}, {Patiri},
  {Weinberg}, \& {Warren}}]{tinker_etal08}
{Tinker}, J.~L., {Conroy}, C., {Norberg}, P., {et~al.} 2008, \apj, 686, 53,
  \dodoi{10.1086/589983}

\bibitem[{{Tinker} {et~al.}(2010){Tinker}, {Robertson}, {Kravtsov}, {Klypin},
  {Warren}, {Yepes}, \& {Gottl{\"o}ber}}]{tinker_etal10}
{Tinker}, J.~L., {Robertson}, B.~E., {Kravtsov}, A.~V., {et~al.} 2010, \apj,
  724, 878, \dodoi{10.1088/0004-637X/724/2/878}

\bibitem[{{van den Bosch} {et~al.}(2013){van den Bosch}, {More}, {Cacciato},
  {Mo}, \& {Yang}}]{van_den_bosch_etal13}
{van den Bosch}, F.~C., {More}, S., {Cacciato}, M., {Mo}, H., \& {Yang}, X.
  2013, \mnras, 430, 725, \dodoi{10.1093/mnras/sts006}

\bibitem[{{van den Bosch} \& {Ogiya}(2018)}]{van_den_bosch18b}
{van den Bosch}, F.~C., \& {Ogiya}, G. 2018, \mnras, 475, 4066,
  \dodoi{10.1093/mnras/sty084}

\bibitem[{{Van Der Walt} {et~al.}(2011){Van Der Walt}, {Colbert}, \&
  {Varoquaux}}]{numpy_ndarray}
{Van Der Walt}, S., {Colbert}, S.~C., \& {Varoquaux}, G. 2011, ArXiv:1102.1523

\bibitem[{{Wang} {et~al.}(2019){Wang}, {Mao}, {Zentner}, {van den Bosch},
  {Lange}, {Schafer}, {Villarreal}, {Hearin}, \& {Campbell}}]{wang_etal19}
{Wang}, K., {Mao}, Y.-Y., {Zentner}, A.~R., {et~al.} 2019, \mnras, 488, 3541,
  \dodoi{10.1093/mnras/stz1733}

\bibitem[{{Watson} {et~al.}(2015){Watson}, {Hearin}, {Berlind}, {Becker},
  {Behroozi}, {Skibba}, {Reyes}, {Zentner}, \& {van den Bosch}}]{watson_etal15}
{Watson}, D.~F., {Hearin}, A.~P., {Berlind}, A.~A., {et~al.} 2015, \mnras, 446,
  651, \dodoi{10.1093/mnras/stu2065}

\bibitem[{{Wechsler} \& {Tinker}(2018)}]{wechsler_tinker_2018}
{Wechsler}, R.~H., \& {Tinker}, J.~L. 2018, \araa, 56, 435,
  \dodoi{10.1146/annurev-astro-081817-051756}

\bibitem[{{White}(2014)}]{white_2014_zeldovich}
{White}, M. 2014, \mnras, 439, 3630, \dodoi{10.1093/mnras/stu209}

\bibitem[{{Wibking} {et~al.}(2020){Wibking}, {Weinberg}, {Salcedo}, {Wu},
  {Singh}, {Rodr{\'\i}guez-Torres}, {Garrison}, \&
  {Eisenstein}}]{wibking_etal20}
{Wibking}, B.~D., {Weinberg}, D.~H., {Salcedo}, A.~N., {et~al.} 2020, \mnras,
  492, 2872, \dodoi{10.1093/mnras/stz3423}

\bibitem[{{Yuan} {et~al.}(2021){Yuan}, {Garrison}, {Hadzhiyska}, {Bose}, \&
  {Eisenstein}}]{yuan_etal21_abacushod}
{Yuan}, S., {Garrison}, L.~H., {Hadzhiyska}, B., {Bose}, S., \& {Eisenstein},
  D.~J. 2021, \mnras, \dodoi{10.1093/mnras/stab3355}

\bibitem[{{Zentner}(2007)}]{zentner_2007_eps_review}
{Zentner}, A.~R. 2007, International Journal of Modern Physics D, 16, 763,
  \dodoi{10.1142/S0218271807010511}

\bibitem[{{Zentner} {et~al.}(2014){Zentner}, {Hearin}, \& {van den
  Bosch}}]{zentner_etal14}
{Zentner}, A.~R., {Hearin}, A.~P., \& {van den Bosch}, F.~C. 2014, \mnras, 443,
  3044, \dodoi{10.1093/mnras/stu1383}

\bibitem[{{Zentner} {et~al.}(2013){Zentner}, {Semboloni}, {Dodelson}, {Eifler},
  {Krause}, \& {Hearin}}]{zentner_etal13}
{Zentner}, A.~R., {Semboloni}, E., {Dodelson}, S., {et~al.} 2013, \prd, 87,
  043509, \dodoi{10.1103/PhysRevD.87.043509}

\bibitem[{{Zhai} {et~al.}(2019){Zhai}, {Tinker}, {Becker}, {DeRose}, {Mao},
  {McClintock}, {McLaughlin}, {Rozo}, \& {Wechsler}}]{zhai_etal19_aemulus3}
{Zhai}, Z., {Tinker}, J.~L., {Becker}, M.~R., {et~al.} 2019, \apj, 874, 95,
  \dodoi{10.3847/1538-4357/ab0d7b}

\bibitem[{{Zheng} \& {Guo}(2016)}]{zheng_guo_2016}
{Zheng}, Z., \& {Guo}, H. 2016, \mnras, 458, 4015, \dodoi{10.1093/mnras/stw523}

\bibitem[{{Zheng} {et~al.}(2005){Zheng}, {Berlind}, {Weinberg}, {Benson},
  {Baugh}, {Cole}, {Dav{\'e}}, {Frenk}, {Katz}, \& {Lacey}}]{zheng_eatl05}
{Zheng}, Z., {Berlind}, A.~A., {Weinberg}, D.~H., {et~al.} 2005, \apj, 633,
  791, \dodoi{10.1086/466510}

\end{thebibliography}


\appendix

\section{Parameterized Subhalo Mass Function}
\label{sec:shmf}

In this section, we describe the fitting function we use to approximate the subhalo mass function in the Bolshoi-Planck simulation \citep[BPL,][]{klypin_etal11}. The BPL simulation was run with cosmological parameters closely matching \citet{planck14b}, and was carried out using the ART code \citep{kravtsov_etal97_art} by evolving $2048^3$ dark-matter particles of mass $m_{\rm p}=1.55\times10^{8}\msun$ on a simulation box of $250\,{\rm Mpc}$ on a side.

For notational convenience, throughout this section we will refer to subhalo mass using the variable $\mh\equiv\log_{10}M_{\rm peak}.$ The basic quantity that we model is the cumulative number density of subhalos as a function of mass, $\Phi_{\rm h}(>\mh).$ The relationship the cumulative mass function and the differential mass function, which we denote by $\phi_{\rm h}(\mh),$ is given by the following equation: 
$$\Phi_{\rm h}(>\mh)\equiv\int_{\mh}^{\infty}\dd \mh' \phi_{\rm h}(\mh').$$

\begin{figure}
\includegraphics[width=8cm]{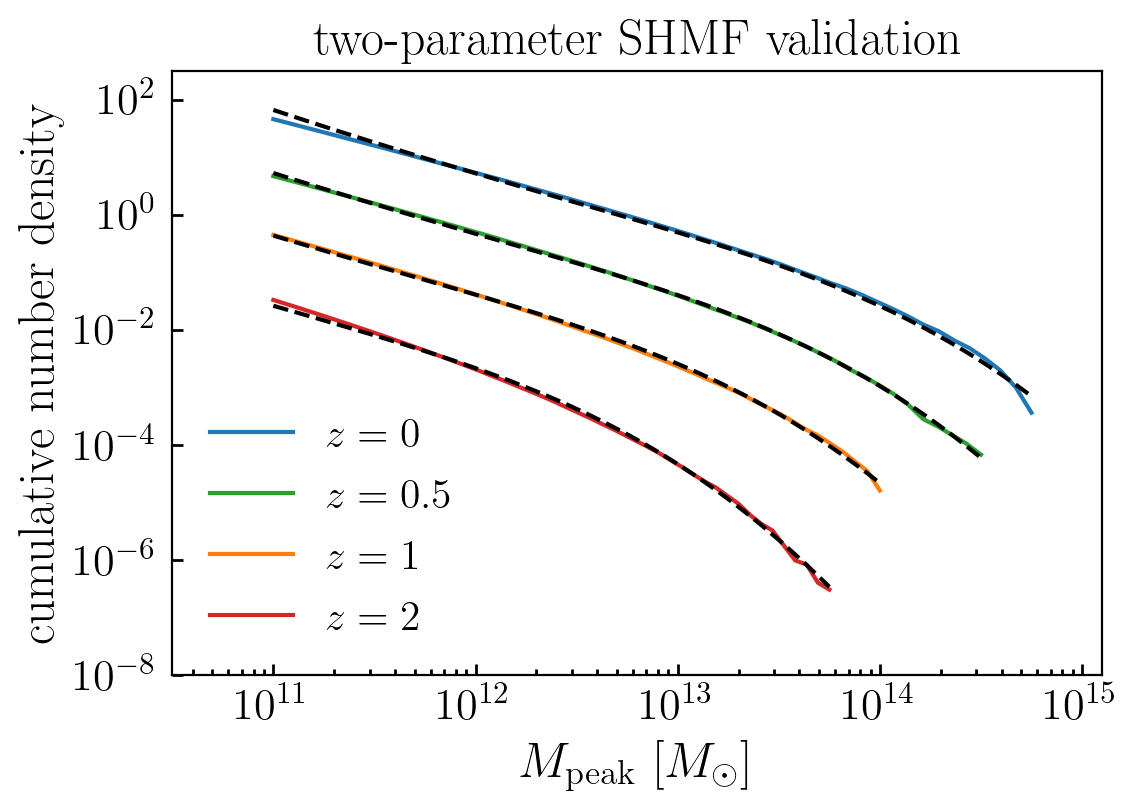}
\caption{Fitting function approximation of the subhalo mass function in the Bolshoi-Planck simulation. Values for $\dd N_{\rm h}(>x)/\dd V$ plotted on the vertical axis are offset by an order of magnitude at different redshifts for visual clarity.}
\label{fig:shmf}
\end{figure}

We parameterize the cumulative mass function to behave like a power-law with constant index at the low-mass end, and like a power-law with a steeply declining power-law index at high mass. We capture this behavior by modeling $\tilde{\Phi}_{\rm h}\equiv\log_{10}\Phi_{\rm h}$ to behave like a linear function of $\mh$ at low mass, and to exhibit a rapidly declining slope above some critical value of subhalo mass, $\mcrit.$ To accomplish this behavior, we utilize a sigmoid function:
\beq
\label{eq:sigmoid}
\\
\mathcal{S}(x\vert\xcrit, k, y_{\rm lo}, y_{\rm hi})\equiv y_{\rm lo} + \frac{y_{\rm hi}-y_{\rm lo}}{1 + \exp(-k\cdot(x-\xcrit))}\nonumber.
\eeq

Using the sigmoid function defined in Eq.~\ref{eq:sigmoid}, we model the cumulative number density as follows:
\beq
\label{eq:sigmoidshmf}
\\
\tilde{\Phi}_{\rm h}(\mh) = y_0 + S(\mh, \mcrit, k, y_{\rm lo}, y_{\rm hi})\cdot(\mh-\mcrit)\nonumber
\eeq
In Eq.~\ref{eq:sigmoidshmf}, the parameter $y_0$ controls the normalization of the subhalo mass function at the critical mass $\mcrit;$ the power-law index at the low-mass end is controlled by $y_{\rm lo},$ and the shape of the high-mass cutoff is characterized by both $k$ and $y_{\rm hi}.$ We note that basic theoretical considerations lead one to expect the true halo mass function to have an exponential cutoff at the high-mass end \citep[see, e.g.,][]{press_schechter_1974,zentner_2007_eps_review}, whereas even in the limit of infinite mass, Eq.~\ref{eq:sigmoidshmf} never attains exponentially declining behavior. For our purposes, however, this shortcoming is immaterial, since our fitting function closely mimics an exponential decline for all halo masses relevant for the present study, and we find that Eq.~\ref{eq:sigmoidshmf} has more numerically stable behavior in applications requiring automatic differentiation at the high-mass end.

We calibrate the values of our fitting function parameters using publicly available\footnote{\url{https://www.peterbehroozi.com/data.html}} subhalo catalog identified with Rockstar and ConsistentTrees \citep{behroozi_etal13_rockstar, behroozi_etal13_consistent_trees, rodriguez_puebla_etal16}. While fitting the free parameters, we hold fixed $y_{\rm lo}=-1.4$, $k=0.9,$ and $y_{\rm hi}=-5.25.$ The remaining two parameters, $y_0$ and $\mcrit,$ exhibit redshift dependence that is well approximated with the sigmoid function defined by Eq.~\ref{eq:sigmoid}. That is, we model $\mcrit(z)$ as:
\beq
\mcrit(z) &=& \mathcal{S}(z, z_{\rm c}^{\mcrit}, k_{\rm c}^{\mcrit}, y_{\rm lo}^{\mcrit}, y_{\rm hi}^{\mcrit}) \nonumber,
\eeq
and similarly for $y_0(z).$ For our best-fitting parameters controlling $\mcrit(z),$ we have $z_{\rm c}^{\mcrit}=0.75, k_{\rm c}^{\mcrit}=1, y_{\rm lo}^{\mcrit}=16.175,$ and $y_{\rm hi}^{\mcrit}=12.925.$ For $y_0(z),$ we have $z_{\rm c}^{y_0}=0.5, k_{\rm c}^{y_0}=1.35, y_{\rm lo}^{y_0}=-8.1,$ and $y_{\rm hi}^{y_0}=-5.9.$ Throughout this paper, we define \shamnet according to the $z=0$ mass function, and we supply these parameters as a convenience for future applications.

In Figure \ref{fig:shmf} we show the fidelity with which our calibration approximates the cumulative mass function of subhalos in BPL; results for mass functions at different redshifts are color-coded as indicated in the legend, and are offset from one another by an order of magnitude for visual clarity. Over the halo mass range $12\leq\mh\leq14,$ the fitting function at $z=0$ is accurate at the $\sim5\%$ level, but at $\mh\approx11$ the accuracy degrades to the 40\% level, and at $\mh\approx14.5$ to the 20\% level. Improving the accuracy of our fitting function at the high-mass end would require a larger simulation than BPL, since the BPL box size of 250 Mpc results in only very few cluster-mass halos. Improving the accuracy at the low-mass end could be accomplished by introducing an additional degree of freedom in the mass-dependence of the power-law index. However, recent work indicates that numerical artifacts and artificial subhalo disruption can result in as large as a factor of two uncertainty in the true SHMF at this mass in a simulation with BPL resolution \citep{van_den_bosch18b,campbell_etal18}, and so efforts to improve the accuracy of such a fitting function at the low-mass end should go hand-in-hand with the development of a galaxy--halo model that has flexibility to capture how subhalo disruption (artificial or otherwise) impacts the galaxy at the subhalo center. We refer the reader to the concluding portion of \S\ref{sec:discussion} for further discussion of how \shamnet would need to be modified in order to derive constraints from observational data.

\section{Analytical SMF predictions}
\label{sec:analytic_smf}

In \S\ref{subsec:diffsmf}, we outlined a technique to make a differentiable prediction for the SMF based on an input subhalo catalog. In this section, we describe a variation on this method that allows one to predict the SMF based an analytical subhalo mass function (SHMF), such as the one supplied in Appendix~\ref{sec:shmf}.

In the presence of scatter in the stellar-to-halo mass relation (SMHM), the relationship between the SMF, $\phi_{\rm g}(\Mstar),$ the SHMF, $\phi_{\rm h}(\mpeak),$ and the SMHM, $\smhmavg,$ is given by Equation ~\ref{eq:shamscatter}, repeated here for convenience:
\beq
\label{eq:shamscatter2}
\\
\phi_{\rm g}(\Mstar) &=& \int_{0}^{\infty}\dd\mpeak\phi_{\rm h}(\mpeak)P(\Mstar\vert\mpeak)\nonumber,
\eeq
where scatter in the SMHM is encoded by $P(\Mstar\vert\mpeak).$ By using the Jacobian of the inverse SMHM, ${\dd\mpeak}/{\dd\Mstar},$ we can change the integration variables of Eq.~\ref{eq:shamscatter2}, so that
\beq
\label{eq:shamscatter2b}
\\
\phi_{\rm g}(\Mstar) &=& \int_{0}^{\infty}\dd\Mstar'\phi_{\rm h}(\mpeak)\frac{\dd\mpeak}{\dd\Mstar'}P(\Mstar\vert\Mstar')\nonumber,
\eeq
where $P(\Mstar\vert\Mstar')$ is a log-normal distribution centered at $\smhmavgprime.$

To differentiably calculate the left-hand side of Equation \ref{eq:shamscatter2b}, one can simply tabulate the integrand on the right-hand side using a grid that is sufficiently broad to cover the support of convolution, and sufficiently dense to achieve the desired precision:
\beq
\label{eq:shamscatter3}
\phi_{\rm g}(\Mstar) =\sum_{i}\Delta\Mstari' && \phi_{\rm h}(\mpeak)\frac{\dd\mpeak}{\dd\Mstari'}\nonumber\\
&& \times P(\Mstar\vert\Mstari').
\eeq
The principal difference between Equation~\ref{eq:shamscatter3} and Equation~\ref{eq:diffsmf} in the main body of the paper is the presence of the Jacobian, ${\dd\mpeak}/{\dd\Mstar};$ this factor does not appear in the simulation-based formulation because its effect is accounted for by the relative abundance of simulated halos as a function of $\mpeak.$

Equation \ref{eq:shamscatter3} makes it plain to see how to calculate derivatives of $\phi_{\rm g}(\Mstar\vert\theta)$ with respect to the model \shamnet parameters, $\theta$: the gradient operator passes through the summation, and $\partial\phi_{\rm g}(\Mstar)/\partial\theta$ can be calculated by simply summing each term. For the analytical approximations we use throughout this paper, each of these terms can in principle be calculated symbolically; we refer the reader to the \href{https://github.com/ArgonneCPAC/shamnet/blob/master/shamnet/smf\_scatter\_convolution.py}{smf\_scatter\_convolution.py} module in our source code for our JAX-based computation of the gradients of Equation \ref{eq:shamscatter3} based on automatic differentiation.
\section{Three-roll stellar-to-halo mass relation}
\label{sec:threeroll}

As discussed in Appendix \S\ref{sec:shamnet_train}, in order to train \shamnet we used an analytical parameterization of the stellar-to-halo mass relation to support the initialization of the network. In \S\ref{subsec:threeroll_def}, we describe the functional form we use for this purpose, and in \S\ref{subsec:threeroll_bestfit} we detail our procedure for optimizing the parameters of this model as a function of the variables $\theta$ that define the abundance matching problem space reviewed in \S\ref{subsec:shamnet_ingredients}.

\subsection{Basic behavior of the three-roll SMHM}
\label{subsec:threeroll_def}

In this section, we describe a new functional form we developed to capture the principal scaling relation of abundance matching, the stellar-to-halo-mass relation (SMHM), i.e., $\smhm.$ In its general shape, the SMHM relation exhibits a characteristic peak at $\mhalo\approx10^{12}\msun,$ and declines roughly like a power-law at both the low- and high-mass ends. We parametrically capture this behavior in terms of the relationship between $\mh\equiv\log_{10}\mpeak,$ and $\ms\equiv\log_{10}\Mstar:$
\beq
\label{eq:rollingplaw}
\ms(\mh) = m_{0} + \gamma(\mh)\cdot(\mh-m_{\rm crit}).
\eeq

From Eq.~\ref{eq:rollingplaw} we see that $\Mstar$ scales like a power-law with $\mhalo,$ with rolling index $\gamma,$ and normalization $m_{0}$ defined by the stellar mass at halo mass $m_{\rm crit}.$  For the function $\gamma(x),$ we use the sigmoid function $\mathcal{S}(x)$ defined in Eq.~\ref{eq:sigmoid} to control the transition between the low-mass slope, $\gamma_{\rm lo}=\gamma(x\rightarrow0),$ and the high-mass slope, $\gamma_{\rm hi}=\gamma(x\rightarrow\infty).$ That is, we have $$\gamma(\mh)=\mathcal{S}(\mh, m_{\rm crit}, k_{\rm crit}, \gamma_{\rm lo}, \gamma_{\rm hi}).$$
We further allow the power-law indices at low- and high-mass to be functions of halo mass, $\gamma_{\rm lo}\rightarrow\gamma_{\rm lo}(\mh),$ and $\gamma_{\rm hi}\rightarrow\gamma_{\rm hi}(\mh),$ where the mass-dependence is again controlled by a sigmoid as defined by Eq.~\ref{eq:sigmoid}. The sigmoid-dependence to $\gamma_{\rm lo}(\mh)$ and $\gamma_{\rm hi}(\mh)$ gives the model freedom for an additional roll in the power-law index at both low- and high-mass ends. 

The above formulation gives our SMHM relation a total of eleven parameters: $m_0, m_{\rm crit}, k_{\rm crit},$ plus four parameters for each of the two sigmoid functions, $\gamma_{\rm lo}(\mh),$ and $\gamma_{\rm hi}(\mh).$ We find that none of the three $k$ parameters provides a physically useful degree of freedom, and so in all of our applications of this SMHM relation, we hold these fixed to unity, giving us an eight-dimensional model for this three-roll SMHM relation. 

\begin{figure}
\includegraphics[width=8cm]{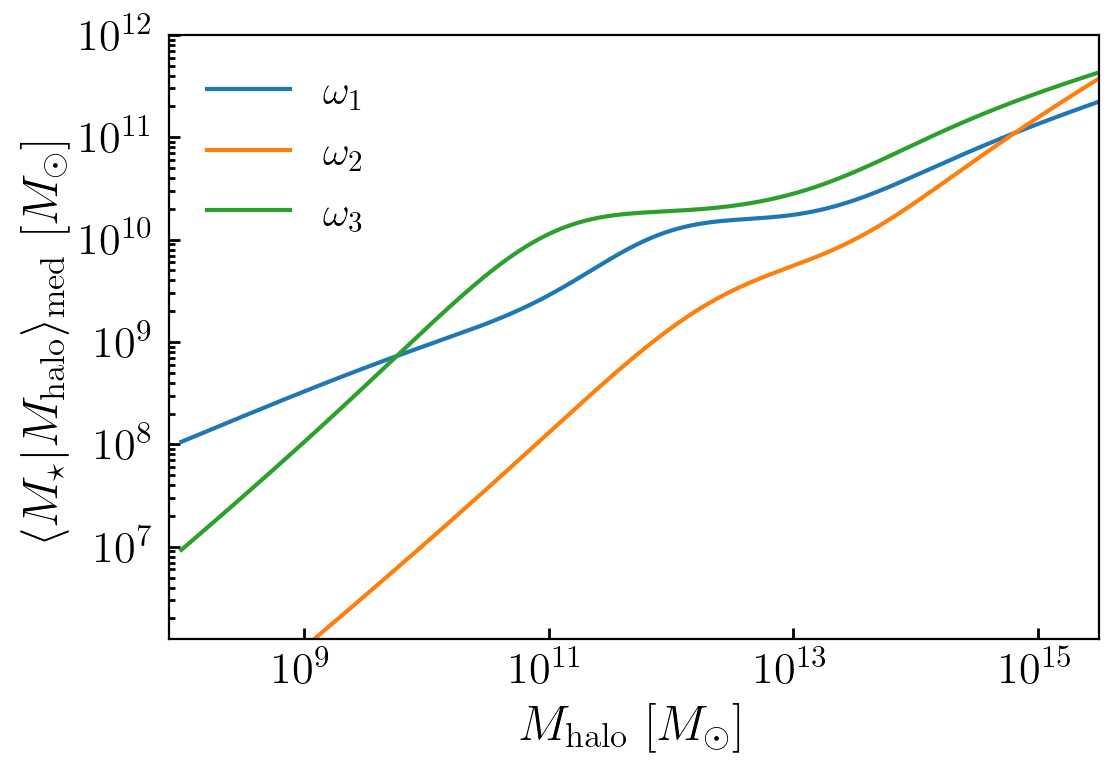}
\caption{{\bf Flexible model for the stellar-to-halo-mass relation (SMHM).} Using the model described in Appendix \S\ref{sec:threeroll},  different curves show the median stellar mass as a function of halo mass for different values of $\omega$ in our eight-dimensional parameter space.}
\label{fig:rollingroller}
\end{figure}

\subsection{Parameterized SHAM with the three-roll SMHM}
\label{subsec:threeroll_bestfit}

In this section, we describe how we use the functional form defined in \S\ref{subsec:threeroll_def} to identify particular solutions to the stellar-to-halo mass relation that defines abundance matching (see Equation \ref{eq:shamscatter}). In particular, here describe how we identify an optimal choice of parameters, $\omega,$ such that the functional form $\ms(\mh\vert\omega)$ defined in Equation \ref{eq:rollingplaw} supplies a solution to Equation \ref{eq:shamscatter} for an input $\theta$ that defines the SMF, $\phi_{\rm g}(\Mstar\vert\theta),$ the SHMF, $\phih(\mpeak\vert\theta),$ and log-normal scatter, $\sigma(\Mstar\vert\mpeak),$ where the $\theta$-dependence of these quantities is specified in \S\ref{sec:shamnet}.

Since the goal of solving Equation \ref{eq:shamscatter} is to define a scaling relation that gives rise to some desired SMF, then in order to solve for $\omega$ we must quantify our choice for the closeness of two stellar mass functions, $\phi_1(\Mstar)$ and $\phi_2(\Mstar).$ Due to the exponentially declining nature of the Schechter function, we have found that minimizing the simple least-squares difference between two SMFs can lead to numerically unstable results during gradient descent with automatic differentiation, particularly when seeking high-quality fits at the very massive end. Computing the logarithmic difference between the two mass functions improves the numerical instability, but does not resolve it; we also find instabilities when implementing a hard-edged clip, $\phigal(\Mstar)\rightarrow{\rm max}\left\{\phigal(\Mstar), \phi_{\rm min}\right\},$ due to the discontinuity of the derivative at the tiny value of the clip. To resolve these issues, we have found it beneficial to use a differentiable clipping function, $\mathcal{R}(x; y_{\rm min}),$ defined as follows
\beq
\label{eq:lupton}
\\
\mathcal{R}(x; y_{\rm min})\equiv\frac{1}{\ln(10)}\left({\rm sinh}^{-1}(x/2y_{\rm min}) + \ln(y_{\rm min})\right)\nonumber,
\eeq
which largely behaves like the base-10 logarithm, but smoothly asymptotes to $y_{\rm min}$ rather than falling below this value \citep[note that this is the same transformation used to define magnitudes of SDSS galaxies in][]{lupton_gunn_szalay99}. Thus when minimizing the difference between two stellar mass functions, in practice we minimize the difference between  $\mathcal{R}(\phi_1(\Mstar), \phi_{\rm min})$ and $\mathcal{R}(\phi_2(\Mstar), \phi_{\rm min}),$ using $\phi_{\rm min}=10^{-15}{\rm Mpc^{-3}M_{\odot}^{-1}}$ to protect against numerical instabilities in the gradient evaluations.

For each choice of $\theta,$ we searched our model parameter space for an optimum value of $\omega$ by minimizing the quantity $\mathcal{L}_{\rm MSE},$ defined as
\beq
\label{eq:smhm_loss}
\mathcal{L}_{\rm MSE}(\theta,\omega) &\equiv& \frac{1}{N}\sum_{i}\left(y^{\rm i}(\theta)-x^{\rm i}(\theta,\omega)\right)^2\nonumber\\
y^{\rm i}(\theta) &=&\mathcal{R}(\phigal(\Mstari\vert\phi_{\ast},\alpha))\\
x^{\rm i}(\theta,\omega)&=&\mathcal{R}(\phigal(\Mstari\vert\theta,\omega))\nonumber
\eeq
where $y^{\rm i}(\theta)$ is the differentiably-clipped logarithm of target Schechter function defined only by $\{\phi_{\ast},\alpha\},$ and $x^{\rm i}(\theta,\omega)$ is the theoretical prediction for the SMF, computed by the method described in Appendix~\ref{sec:analytic_smf}. We calculate $\mathcal{L}_{\rm MSE}$ by evaluating the predicted and target mass functions at a set of $N$ control points, $\Mstar^{\rm i},$ using $N=100$ logarithmically-spaced values spanning the range $10^{9}\msun<\Mstar<10^{12}\msun.$ To minimize $\mathcal{L}_{\rm MSE}$, we use the JAX implementation of the Adam algorithm \citep{kingma_ba_adam_2015}, which is a gradient descent technique with an adaptive learning rate, where we also use JAX to compute the gradients. We use 2 successive burn-in cycles with a step-size parameter $s=0.05$ for $\sim50$ updates, followed by $1000$ updates with $s=0.01.$

In this minimization calculation, the eight parameters $\omega$ defined in \S\ref{subsec:threeroll_def} are not the actual variables we use to calculate our predicted values for the SMF; instead we minimize $\omega',$ defined by:$$\omega'\equiv\mathcal{S}(\omega\vert\omega_0,k_{\omega},\omega_{\rm lo},\omega_{\rm hi}),$$ where $\mathcal{S}(x)$ is the sigmoid function defined by Eq.~\ref{eq:sigmoid}. We tailor the lower and upper bounds of each dimension of $\omega$ according to reasonable expectations for physically plausible SMHMs, and we hold $k_{\omega}$ fixed to a constant value of $0.1,$ and $\omega_0$ fixed to the halfway point between $\omega_{\rm lo}$ and $\omega_{\rm hi}.$ We refer the reader to \href{https://github.com/ArgonneCPAC/shamnet/blob/master/shamnet/shamnet_traindata.py}{the relevant section of our source code} for additional details of our implementation.

\section{SHAMNet Definition and Training}
\label{sec:shamnet_train}

In this appendix, we describe our procedure for training \shamnet, $\fsham(\mpeak,\theta\vert\psi).$ We remind the reader that in this notation, the parameters $\psi$ refer to the weights and biases of the neural network described in detail in this section, and as described in \S\ref{subsec:shamnet_ingredients}, the parameters $\theta$ control the behavior of the three ingredients needed to define SHAM:
\begin{enumerate}
\item $\phi_{\rm g}(\Mstar\vert\theta),$ the galaxy stellar mass function,
\item $\phih(\mpeak\vert\theta),$ the subhalo mass function described in \S\ref{sec:shmf},
\item $\sigma(\mpeak\vert\theta),$ the halo mass-dependent scatter.
\end{enumerate}

As outlined in \S\ref{sec:sham}, for a particular choice of $\theta,$ SHAM is defined to be the scaling relation, $\smhmtheta,$ that gives rise to the target stellar mass function, $\phi_{\rm g}(\Mstar\vert\theta),$ when applied to the subhalo population, $\phi_{\rm h}(\mpeak\vert\theta),$ in the presence of (log-normal) scatter, $\sigma(\mpeak\vert\theta).$ The goal of training \shamnet is to identify the parameters $\psi$ such that for all physically relevant values of $\theta,$ the function $\fsham(\mpeak,\theta\vert\psi)=\smhmtheta$ satisfies the SHAM-defined relationship between $\phigal,$ $\phih,$ and $\sigma.$  The nature of this objective makes training $\fsham(\mpeak,\theta\vert\psi)$ different from a standard problem in neural network regression, since there does not exist a target function whose output defines the training data; instead, the parameters $\psi$ will be considered optimal when the SMF that emerges from $\smhmtheta$ agrees with $\phi_{\rm g}(\Mstar\vert\theta)$ to the desired precision.

\begin{table}[h!]
\label{table:training}
  \begin{center}
    \caption{Summary of two-phase training of \shamnet.}
    \label{tab:table1}
    \begin{tabular}{l|c|r} 
      \textbf{Training Phase} & \textbf{Target data} & \textbf{Loss function}\\
      \hline
      initialization & $\smhmavg,$ Eq.~\ref{eq:rollingplaw} & Eq.~\ref{eq:initializer_loss}\\
      final & $\phi_{\rm g}(\Mstar)$, Eqs.~\ref{eq:shamscatter2}-\ref{eq:shamscatter3} & Eq.~\ref{eq:shamnet_loss}\\
    \end{tabular}
  \end{center}
\end{table}

We begin in \S\ref{subsec:shamnet_architecture} with a description of the design of the neural network we use for $\fsham.$ Our training then proceeds in two phases. In the first phase described in \S\ref{subsec:shamnet_init}, we tune the network parameters $\psi$ so that $\fsham(\mpeak,\theta\vert\psi)$ closely agrees with a parametric form for the stellar-to-halo mass relation that has been tuned in advance to approximately solve Eq.~\ref{eq:shamscatter}. After this initial phase, we optimize the parameters $\psi$ using the training procedure described in \S\ref{subsec:shamnet_training}. We will provide a reasonably comprehensive description of this two-phase procedure in this appendix, but we refer the reader to \url{https://github.com/ArgonneCPAC/shamnet} for all the quotidian details. A summary of this two-phase training appears in the table below.

\subsection{Architecture}
\label{subsec:shamnet_architecture}

In this section, we describe the architecture of the neural network we use for \shamnet, which provides a mapping
$$\fsham:\{\theta,\mpeak\}\rightarrow\Mstar$$ based on a simply-connected multi-layer perceptron (MLP). For the input variables $\theta,$ we use a $2$-parameter Schechter function to capture variations in the galaxy stellar mass function, and a 2-parameter sigmoid to describe the halo mass dependence of scatter, adopting the forms described in \S\ref{subsec:shamnet_ingredients}. For the subhalo mass function, we use the parametric model defined in \S\ref{sec:shmf}, but in this case we hold these parameters fixed to the values calibrated to approximate the $z=0$ mass function in the Bolshoi simulation.  Thus in addition to $\mpeak,$ \shamnet accepts a 5-dimensional parameter, $\theta,$ and returns stellar mass, $\Mstar,$ so that $\fsham:\mathbb{R}^5\rightarrow\mathbb{R}^1.$

Our neural network $\fsham$ thus has an input layer of 5 nodes, one for each dimension of $\{\theta,\mpeak\},$ and an output layer of a single node, $\Mstar.$ Between the inputs and outputs, we use 4 hidden layers composed of $64, 32, 8,$ and $4$ nodes, respectively. Each node in a layer is connected to every node in the previous layer, so that all 4 of our hidden layers are dense. As with all simply-connected MLPs, the computations performed by the $i^{\rm th}$ node of a layer, $y_{\rm i},$ consists of two successive operations performed on the outputs of all the nodes of the previous layer, $x_{\rm j}.$ The first set of operations consists of a linear transformation,
\beq
\label{eq:nnlinear}
y_{\rm i}= \sum_{\rm j}W_{\rm ij}x_{\rm j} + b_{\rm i},
\eeq
and the second operation is a nonlinear transformation, $\activation,$ that is independently applied to the each node $y_{\rm i}$ in the layer; the result $\activation(y_{\rm i})$ is then fed as an input to each node in the next layer. For the activation function $\activation$ in every node of each dense hidden layer, we use a SELU function,
\beq
\label{eq:selu}
\activation(z) = \lambda\begin{cases} z &\mbox{if } z > 0 \\
\alpha e^z-\alpha & \mbox{if } z \leq 0 \end{cases},
\eeq
where $\lambda=1.05$ and $\alpha=1.67$ are chosen so that the mean and variance of the inputs are preserved between two consecutive layers \citep{klambauer_etal17}. For the activation function applied to the output layer, we use the sigmoid function defined in Eq.~\ref{eq:sigmoid} to enforce that the returned stellar mass is bounded within the range $10^{-5}\msun<M_{\star}<10^{25}\msun.$ Our MLP is implemented in the {\tt stax} subpackage of the JAX library.

\subsection{Initialization}
\label{subsec:shamnet_init}

Our goal for training \shamnet is to identify the weights, $W_{\rm ij},$ and biases, $b_{\rm j},$ collectively represented with the variable $\psi,$ such that $\fsham$ provides the desired mapping from $\mpeak\rightarrow\Mstar$ for any physically relevant parameters, $\theta.$ We begin our procedure for optimizing $\psi$ by generating approximations to the desired mapping, $\smhmtheta,$ for a large collection of $\theta;$ the collection of $\theta$ and associated mappings will serve as training data for the first phase of optimizing $\fsham.$ Thus at the end of this phase, we will have a collection of weights and biases, $\psi_{\rm init},$ that roughly provides the desired scaling relation, $\smhmtheta,$ for each $\theta$ in the training set. The point in parameter space $\psi_{\rm init}$ will be used as the starting point for the optimization procedure described in \S\ref{subsec:shamnet_training}, and so we refer to calculations described in the present section as the initialization phase of our training.

As described in \S\ref{sec:shamnet}, the parameters $\theta$ define a specific combination of $\phigal(\Mstar\vert\theta), \phih(\mpeak\vert\theta),$ and $\sigma(\mpeak\vert\theta);$ for each such combination, we seek to determine the abundance matching relation, $\smhm,$ that provides a solution to Eq.~\ref{eq:shamscatter}. In the initialization phase of our training, we assume that the parametric form $\ms(\mh\vert\omega),$ as defined by Eq.~\ref{eq:rollingplaw}, is sufficiently flexible to supply an adequate stellar-to-halo mass relation, and we train \shamnet to reproduce this parameterized relation. We will use the notation $\omega_{\theta}$ to refer to the set of parameters of the three-roll SMHM that minimize Eq.~\ref{eq:smhm_loss} using the optimization techniques detailed in Appendix~\ref{subsec:threeroll_bestfit}. Thus in order to generate training data for the initialization phase of our training, we need to generate a large collection of pairs, $\{\theta, \omega_{\theta}\},$ where the values of $\theta$ span the physically relevant range, and the accompanying value $\omega_{\theta}$ represents an approximate solution to Eq.~\ref{eq:shamscatter}. To generate the collection of pairs, $\{\theta, \omega_{\theta}\},$ we randomly draw a value of $\theta$ from an observationally relevant range, and for each random draw, we use the gradient descent technique described in Appendix~\ref{sec:threeroll} to find a point $\omega_{\theta}$ that optimally solves Eq.~\ref{eq:shamscatter}. 

Since $\theta=\{\phi_{\ast},\alpha,y^{\sigma}_{\rm lo}, y^{\sigma}_{\rm hi}\},$ then we have four dimensions for which we need to define an observationally relevant range. The last two dimensions control the level of $\mhalo$-dependent scatter, which we allow to vary independently within the open interval, (0.1, 0.5). The first two variables define the variations in the SMF, which for purposes of this paper, we wish to be reasonably close to SDSS measurements of the low-redshift universe \citep[e.g.,][]{li_white_2009}. To determine the relevant range of the SMF parameters, we define an SDSS-like SMF, $\phi_{\rm SDSS}(\Mstar),$ by selecting fiducial values $\phi_{\ast}=0.005\ {\rm Mpc^{-3}\msun^{-1}},$ and $\alpha=-1.06.$ Using {\tt emcee} \citep{emcee_2019}, we run an MCMC to determine posteriors on the parameters $\phi_{\ast}$ and $\alpha,$ assuming a Gaussian likelihood with a diagonal covariance matrix defined by $\Mstar$-independent uncertainty of $0.1$ dex on $\phi_{\rm SDSS}(\Mstar).$ The results of this MCMC supply a Gaussian-like distribution of points in our SMF parameter space, and we fit this posterior distribution with a two-dimensional Gaussian distribution, $\mathcal{N}(\mu, {\rm Cov(}\phi_{\ast},\alpha)).$ When generating training data for \shamnet, we sample SMF parameters by randomly draw pairs $\{\phi_{\ast},\alpha\}$ based on a Latin Hypercube, where the axes of the Latin Hypercube are aligned with the eigenvectors of ${\rm Cov(}\phi_{\ast},\alpha),$ and span a $5\sigma$ length in each direction; any values with $\alpha<1$ correspond to non-monotonic SMFs, and are discarded.\footnote{We use the {\tt pyDOE2} package\footnote{\url{https://github.com/clicumu/pyDOE2}} in our random sampling based on a Latin Hypercube \citep{mckay_beckman_conover_1979_latin_hypercubes,iman_helton_campbell_1981_latin_hypercubes}.}

For each value of $\theta$ generated by this sampling method, we identify a best-fitting stellar-to-halo mass relation, $\omega_{\theta},$ using the optimization techniques described in \S\ref{subsec:threeroll_bestfit}; the quality of each fit is quantified by $\mathcal{L}_{\rm MSE}(\omega,\theta),$ as defined in Eq.~\ref{eq:smhm_loss}. For some points $\theta,$ the best-fitting value $\omega_{\theta}$ produces only a rough recovery of the target stellar mass function, $\phigal(\Mstar\vert\theta).$ In the initialization phase of training \shamnet, we discard all pairs $\{\theta, \omega_{\theta}\}$ with $\mathcal{L}_{\rm MSE}>0.05,$ which we find is about 5\% of the sampling data. While this rejection rate implies that our training set will not fairly sample the full, observationally relevant region of parameter space as defined above, this is rather harmless because the primary purpose of this procedure is just to generate {\em some} collection $\{\theta, \omega_{\theta}\}$ used to identify $\psi_{\rm init}.$ As described in \S\ref{subsec:shamnet_architecture}, the design of \shamnet has no reliance upon the functional form of the stellar-to-halo mass relation defined in Appendix~\ref{sec:threeroll}, $\ms(\mh\vert\omega)$, and so rejected points $\omega_{\theta}$ simply represent an inflexibility of the three-roll SMHM functional form, but this will not impact \shamnet as we will not reject such pairs $\{\theta, \omega_{\theta}\}$ after $\psi_{\rm init}$ has been identified.

The goal of the initialization phase of our training is to identify a set of weights and biases, $\psi_{\rm init},$ such that the behavior of our neural network, $\fsham,$ gives an accurate approximation of the parametric function defined in Appendix \ref{sec:threeroll}. To achieve this goal, we seek to identify the point $\psi_{\rm init}$ that minimizes the following cost function:
\beq
\label{eq:initializer_loss}
\mathcal{L}_{\rm MSE}(\mh,\theta\vert\psi) &\equiv& \left(y(\mh,\theta)-x(\theta\vert\psi)\right)^2\nonumber\\
y(\mh,\theta) &=& \ms(\mh\vert\omega_{\theta})\\
x(\theta\vert\psi)&=& \fsham(\mh,\theta\vert\psi)\nonumber
\eeq
In evaluating $\mathcal{L}_{\rm MSE},$ we used batch sizes of 50 pairs $\{\theta, \omega_{\theta}\},$ at a time, sampling values of $\theta$ using the procedure outlined above; for each such pair, we evaluated the predicted and target value of $\ms$ at 200 points in halo mass, using uniform random sampling in $\mh$ spanning the range $9<\mh<16.$ We used the implementation of the Adam optimizer in JAX with a step-size parameter of 0.001 to identify $\psi_{\rm init}$ through 1500 gradient evaluations, which we found through experimentation to be sufficient for the network to achieve the level of accuracy required for this initializing phase of training the network.

\subsection{\shamnet Training}
\label{subsec:shamnet_training}

The goal of the final phase of our training is to identify a set of weights and biases, $\psisham,$ such that the behavior of our neural network, $\fsham,$ supplies a stellar-to-halo mass relation that accurately solves the defining equation of abundance matching, Eq.~\ref{eq:shamscatter}. We note that if the parametric function $\ms(\mh\vert\omega)$ defined in Appendix \ref{sec:threeroll} were sufficiently flexible, then this goal would already be achieved via the identification of $\psi_{\rm init}.$ However, as described in Appendix~\ref{subsec:shamnet_init}, roughly 5\% of points in the observationally relevant range have SMFs that cannot be described to high accuracy using a stellar-to-halo mass relation defined by this functional form. Thus in the final training phase of \shamnet, we no longer rely on training data based on the parametric function $\ms(\mh\vert\omega).$ Instead, we seek to identify the point $\psisham$ that directly minimizes the difference between the predicted and target SMF. Our cost function in the final phase of training is therefore defined by minimizing the mean squared error between the predicted and target SMF across the desired range of stellar mass bins:
\beq
\label{eq:shamnet_loss}
\mathcal{L}_{\rm MSE}(\theta\vert\psi) &\equiv& \frac{1}{N_{\rm bins}}\sum_{i}\left(y^{\rm i}(\theta)-x^{\rm i}(\theta)\right)^2\nonumber\\
y^{\rm i}(\theta) &=&\mathcal{R}(\phigal(\Mstari\vert\phi_{\ast},\alpha))\\
x^{\rm i}(\theta\vert\psi)&=&\mathcal{R}(\phigal(\Mstari\vert\theta,\psi))\nonumber.
\eeq
In Eq.~\ref{eq:shamnet_loss}, the values of $y^{\rm i}(\theta)$ are determined by the target Schechter SMF. The predicted values of the SMF, $x^{\rm i}(\theta\vert\psi),$ are derived using the same techniques detailed in Appendix~\ref{sec:analytic_smf} for calculating the SMF in the presence of scatter via an analytical expression for the SMHM, where in this case the analytical SMHM is given by $\fsham(\mh,\theta\vert\psi).$

We use the Adam optimizer with a step size parameter of 0.001 to minimize $\mathcal{L}_{\rm MSE}(\theta),$ where each of our $10^5$ cost function evaluations used a batch size of $N=50$ points that span the parameter space of $\theta,$ again using the same sampling method detailed in \S\ref{subsec:shamnet_init}. The end result of this procedure is the identification of $\psisham,$ which defines the behavior of the $\fsham$ function whose performance is illustrated in Figure~\ref{fig:shamnet_accuracy}.

\section{Triweight Kernel Convolutions}
\label{sec:tdubs}

\begin{figure}
\includegraphics[width=8cm]{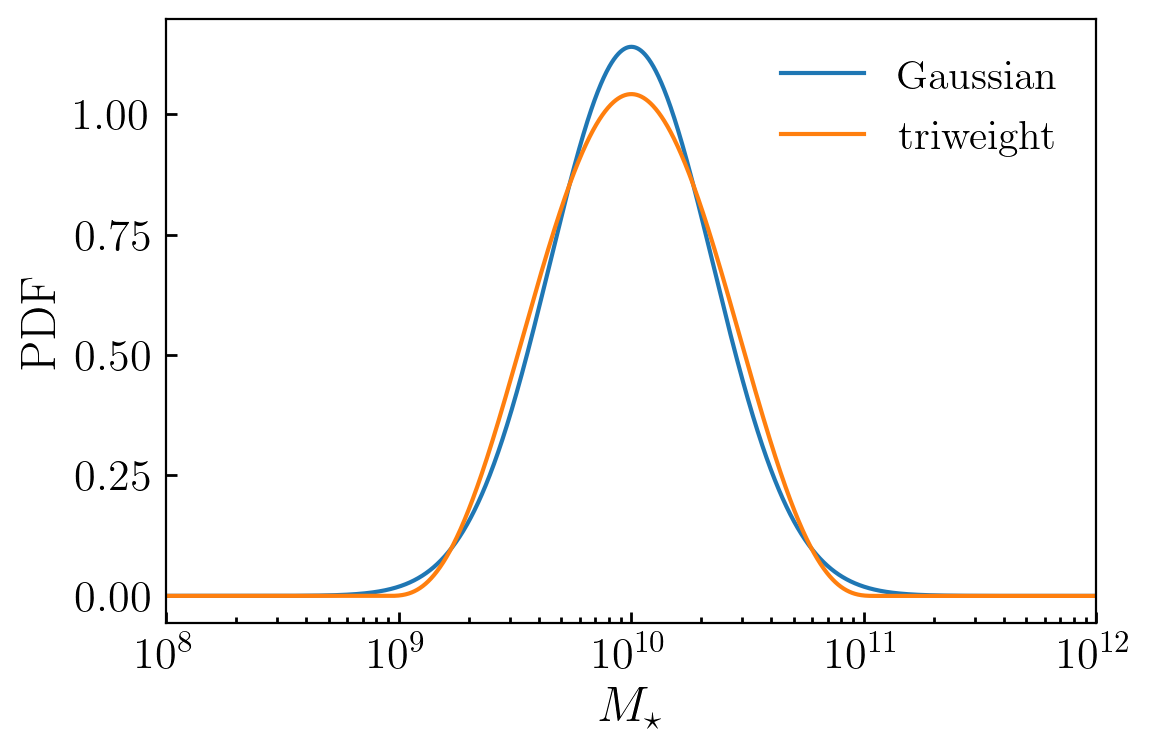}
\caption{{\bf Triweight convolution kernel}. The blue curve shows the probability density function of a typical log-normal distribution in stellar mass, $\mathcal{N}(\mu=10, \sigma=0.25);$ the orange curve shows the triweight kernel, $\mathcal{K},$ defined in Eq.~\ref{eq:triweight}. The triweight function $\mathcal{K}(x)$ vanishes at points beyond $\mu\pm3\sigma,$ is differentiable for all $x,$ and is highly performant on GPUs.}
\label{fig:triweight}
\end{figure}

Many of the calculations in this paper involve convolutions of log-normal distributions such as the one appearing in Eq.~\ref{eq:shamscatter}. In this section, we review an alternative to Gaussian convolution based on the triweight kernel, $\mathcal{K},$ defined as:
\beq
\label{eq:triweight}
\mathcal{K}(z) \equiv\begin{cases}
	\frac{35}{96}\left[ 1-(z/3)^{2} \right]^{3}, & -3\sigma\leq x\leq3\sigma \\
	0, & \text{otherwise}
\end{cases}
\eeq where $z\equiv(x-\mu)/\sigma.$
The comparison to a Gaussian is shown in Figure~\ref{fig:triweight}. The two distributions have the same first and second moments, $\mu$ and $\sigma,$ but differ in their higher-order moments.

The function $\mathcal{K}(x)$ has several properties that make it convenient for our purposes. As described in \S\ref{sec:diffobs}, differentiable predictions for cosmological summary statistics involve weights, $w_{\rm halo},$ computed by integrating some assumed PDF (typically a log-normal) across some bin(s) of stellar mass. Once $w_{\rm halo}$ has been computed for every simulated subhalo, various point-estimators are applied to the population of synthetic galaxies in order to make predictions for summary statistics.

Because Gaussians are everywhere non-zero, computing weighted two-point summary statistics using exact log-normals can be expensive due to the need to keep track of a large number of tiny contributions. One simple technique to mitigate this computational expense is to impose a clip at some multiple of $\sigma.$ This gains back performance at the cost of some (controllably small) roundoff error, but results in discontinuity in the weights at $x=\mu\pm3\sigma.$ On the other hand, the triweight kernel is a $C^{\infty}$ function on the real line, and points with $\vert x-\mu\vert>\pm3\sigma$ contribute formally zero weight, and so can be neglected from two-point calculations without consequence. Moreover, computations of predictions based on Gaussians require special-function evaluations that can be far slower on GPU accelerator devices in comparison to the small number of elementary arithmetical operations required to evaluate Eq.~\ref{eq:triweight}.

In computing predictions based on $\mathcal{K}$ rather than $\mathcal{N},$ we are, in effect, proposing a slightly different probability distribution $P(\Mstar\vert\mhalo)$ than the traditional log-normal.
However, the difference between models distinguished only by $\mathcal{K}$ and $\mathcal{N}$ is likely to be observationally immaterial, as it is already challenging to obtain tight constraints on the second moment $P(\Mstar\vert\mhalo),$ which is the same in the two distributions.

\section{Simulation-Based $\Delta\Sigma$ Predictions}
\label{sec:simlensing}

In this section, we derive the equations associated with our calculation of $\Delta\Sigma(R),$ the excess surface mass density profile of a sample of points in a cosmological simulation. We highlight that the computation described below is exact, can be used to calculate the lensing of either galaxies or halos simulated with or without hydrodynamics, and applies on all spatial scales resolved by the simulation, including the deeply nonlinear regime. Our approach also facilitates a highly convenient pre-computation of the lensing profile on a per-object basis, so that galaxy-halo model predictions for the lensing of a stacked sample can be calculated simply as a masked or weighted sum over the pre-computed profiles.

We begin by considering the gravitational lensing at a projected distance $R$ from a single point mass, $\mpp.$ The surface density profile, $\Sigma(R),$ satisfies:
\beq
\label{eq:sigmapoint}\nonumber
\int_{R_{\rm min}}^{R_{\rm max}}\Sigma(R)2\pi R{\rm d}R=
	 \begin{cases}
	 	\mpp, & \text{if}\ R_{\rm min} < R < R_{\rm max} \\
		0, & \text{otherwise}
	 \end{cases}
\eeq
Using this expression, we can calculate $\langle\Sigma(R)\rangle,$ the average surface density in an annulus with inner radius $R_{\rm min}$ and outer radius $R_{\rm max}:$
\beq
\label{eq:sigmaann}
\langle\Sigma(R)\rangle &\equiv &\frac{\int_{R_{\rm min}}^{R_{\rm max}}\Sigma(R)2\pi R{\rm d}R}{\int_{R_{\rm min}}^{R_{\rm max}}2\pi R{\rm d}R}\nonumber\\
	&=&
	 \begin{cases}
	 	\mpp/A_{\rm ann}, & \text{if}\ R_{\rm min} < R < R_{\rm max} \\
		0, & \text{otherwise}
	 \end{cases}
\eeq
where $A_{\rm ann}=\pi(R_{\rm max}^{2} - R_{\rm min}^{2})$ is the area of the annulus. 

We can similarly compute $\bar{\Sigma}(<R)=\mpp/\pi R^{2}$ averaged over the same annulus:
\beq
\langle\bar{\Sigma}(<R)\rangle &\equiv & \frac{1}{A_{\rm ann}}\int_{R_{\rm min}}^{R_{\rm max}}\bar{\Sigma}(<R)2\pi R{\rm d}R,\nonumber
\eeq
which reduces to 
\beq
\label{eq:sigmacyl0}
\langle\bar{\Sigma}(<R)\rangle = \frac{2\mpp}{A_{\rm ann}}\ln(R_{\rm max}/a),
\eeq
where the value of $a$ depends on whether the point mass is located inside the annulus:
\beq
\label{eq:sigmacyl1}
	 a=\begin{cases}
		R, & R > R_{\rm min}\\
	 	 R_{\rm min}, & R < R_{\rm min}
	 \end{cases}
\eeq

Using Eqs.~\ref{eq:sigmaann} \& \ref{eq:sigmacyl0} together with Eq.~\ref{eq:dsdef}, we can calculate the value of $\Delta\Sigma$ averaged over the annulus:
\beq
\label{eq:dspoint}
\langle\ds(R)\rangle &\equiv&\langle\bar{\Sigma}(<R)-\Sigma(R)\rangle \\
&=& \frac{\mpp}{A_{\rm ann}}G(R, R_{\rm min}, R_{\rm max}), \nonumber
\eeq
where 
\beq
G&=&\nonumber \begin{cases}
		2\ln(R_{\rm max}/R_{\rm min}), & R < R_{\rm min}\\
		2\ln(R_{\rm max}/R)-1, & R_{\rm min}< R< R_{\rm max}\\
		0, & R_{\rm max}<=R.
	\end{cases}
\eeq
In order to calculate the average lensing profile $\langle\ds(R)\rangle$ about an individual point in a N-body simulation, Equation \ref{eq:dspoint} generalizes in the obvious way as a sum over the point-masses used to trace the simulated density field. For a simulated snapshot at redshift $z_{\rm snap},$ if the cosmological distance $D(z_{\rm snap})$ satisfies  $D(z_{\rm snap})\gg R_{\rm max},$ then we can use any of the Cartesian axes of the snapshot as the line-of-sight, and the remaining two axes to define $R.$ 

For a fixed choice of $R_{\rm min}$ and $R_{\rm max},$ the average lensing profile $\langle\ds\rangle$ can be computed once and for all about every subhalo in a simulated snapshot. Once tabulated, predicting the lensing produced by a stack of subhalos can be computed simply by averaging over the pre-computed values of $\langle\ds\rangle$ for each member of the stack (or alternatively, by calculating an average that has been weighted by a continuously-valued sample selection function, as in \S\ref{subsec:diffds}).


\end{document}